\documentclass[]
{jpp}

\usepackage{natbib}
\usepackage{graphicx}
\usepackage{epstopdf, epsfig}
\usepackage[utf8]{inputenc}
\usepackage[T1]{fontenc}
\usepackage{amsmath}
\usepackage[usenames,dvipsnames]{color}
\usepackage{float}
\usepackage{mathtools}
\usepackage{amsmath,amssymb,amsbsy}
\usepackage{bbm,bm,mathrsfs,yfonts}
\usepackage[capitalize]{cleveref}
\usepackage{xcolor}
\usepackage{microtype}
\usepackage{soul,xcolor}
\setstcolor{red}
\newcommand{\be}{\begin{equation}}
\newcommand{\ee}{\end{equation}}

\renewcommand{\l }{\left}
\renewcommand{\r }{\right}

\DeclarePairedDelimiter\floor{\lfloor}{\rfloor}
\newcommand{\tj}[6]{ \begin{pmatrix}
   #1 & #2 & #3 \\
   #4 & #5 & #6 
\end{pmatrix}}
\usepackage{relsize}

\shorttitle{Nonlinear gyrokinetic Coulomb collision operator}
\shortauthor{R. Jorge, B. J. Frei and P. Ricci}

\title{Nonlinear Gyrokinetic Coulomb Collision Operator}

\author{
  R. Jorge\aff{1,2}\corresp{\email{rjorge@umd.edu}}\thanks{Currently present at Institute for Research in Electronics and Applied Physics, University of Maryland, College Park MD 20742, USA},
  B. J. Frei\aff{1}
  \and P. Ricci\aff{1}
}

\affiliation{
\aff{1}École Polytechnique Fédérale de Lausanne (EPFL), Swiss Plasma Center (SPC), CH-1015
Lausanne, Switzerland
\aff{2}Instituto de Plasmas e Fusão Nuclear, Instituto Superior Técnico, Universidade de Lisboa, 1049-001 Lisboa, Portugal
}

\begin{document}

\maketitle

\begin{abstract}
A gyrokinetic Coulomb collision operator is derived, which is particularly useful to describe the plasma dynamics at the periphery region of magnetic confinement fusion devices.
The derived operator is able to describe collisions occurring in distribution functions arbitrarily far from equilibrium with variations on spatial scales at and below the particle Larmor radius.
A multipole expansion of the Rosenbluth potentials is used in order to derive the dependence of the full Coulomb collision operator on the particle gyroangle.
The full Coulomb collision operator is then expressed in gyrocentre phase-space coordinates, and a closed formula for its gyroaverage in terms of the moments of the gyrocenter distribution function in a form ready to be numerically implemented is provided.
Furthermore, the collision operator is projected onto a Hermite-Laguerre velocity space polynomial basis and expansions in the small electron-to-ion mass ratio are provided.
\end{abstract}


\section{Introduction}

%

The plasma periphery, which encompasses the edge and the scrape-off layer regions, plays a central role in determining the overall performance of a fusion device, as it regulates the overall plasma confinement, it controls the plasma-wall interactions, it is responsible for power exhaust, and it governs the plasma refueling and the removal of fusion ashes \citep{Ricci2015}.
Understanding the plasma dynamics in the periphery is therefore crucial for the success of the whole fusion program \citep{Connor1998}.

While the plasma dynamics in the scrape-off layer has been described mainly using drift-reduced fluid models valid at low frequencies compared to the ion cyclotron frequency, $\omega \ll \Omega_i$, and in the limits $k_\parallel \lambda_{mfpe} \ll 1$ and $k_\perp \rho_i \ll 1$, i.e. short electron mean free path in comparison to the parallel wavelength and {long} perpendicular wavelengths with respect to the Larmor radius \citep{Dudson2009,Tamain2009,Ricci2012a,Halpern2016a,Stegmeir2016,Zhu2018,Paruta2018}, these approximations are often marginal near the separatrix and inside it, i.e. in the edge region.
In fact, even though turbulence is still dominated by low-frequency fluctuations, the plasma in the edge is hotter and less collisional than in the scrape-off layer and the use of a fluid model becomes questionable.
Moreover, in the edge region, small scale $k_\perp \rho_i \sim 1$ fluctuations are important \citep{Hahm2009}.
This is especially relevant in the high-temperature tokamak H-mode regime \citep{Zweben2007}, the regime of operation relevant for ITER and future devices.
Despite recent progress \citep{Chang2017,Shi2017,Pan2018}, overcoming the limitation of the drift-reduced fluid models in the description of the tokamak periphery region by using a gyrokinetic model valid at $k_\perp \rho_i \sim 1$ has proven to be exceptionally demanding.
Among the numerous challenges, the effort is undermined by the lack of a proper collisional gyrokinetic model for the periphery.

%
In fact, with respect to the core, due to the lower temperature values and associated high collisionality, the use of a gyrokinetic model to simulate the plasma dynamics in the tokamak periphery requires an accurate collision operator.
This is necessary as collisions set the level of neoclassical transport and strongly influence the turbulent dynamics by affecting the linear growth rate and nonlinear evolution of turbulent modes \citep{Barnes2009}.

Since the first formulations of the gyrokinetic theory, there have been significant research efforts to take collisions into account \citep{Catto1977,McCoy1981,Start2002,Brizard2004,Abel2008a,Barnes2009,Li2011,Dorf2012,Esteve2015,Hakim2019,Pan2019}.
The first effort devoted to a gyrokinetic collision operator can be traced back to the work of \citet{Catto1977}, later improved by \citet{Abel2008} by adding the necessary terms needed to ensure non-negative entropy production.
The result of this effort is a linearized gyrokinetic collision operator that contains pitch-angle scattering effects and retains important conservation properties.
A linearized gyrokinetic Coulomb collision operator derived from first principles was then presented in \citet{Li2011} and \citet{Madsen2013a}.
However, as turbulence in the tokamak periphery is essentially nonlinear, the relative level of fluctuations in this region being of order unity \citep{Scott2002}, and the level of collisions is not sufficient for a local thermalization, the distribution function may significantly deviate from a local Maxwellian distribution \citep{Tskhakaya2012}.
Therefore, a nonlinear formulation of the gyrokinetic Coulomb collision operator is crucial to adequately describe the dynamics in the periphery.
Only recently, several theoretical studies have emerged in order to derive non-linearized collisional gyrokinetic operators that keep conservation laws in their differential form.
In particular, we mention the recent Poisson bracket formulations of the full nonlinear Coulomb collision operator \citep{Brizard2004,Sugama2015,Burby2015}.
While the formulation of these operators represent significant progress, the presence of a six-dimensional phase-space integral in these expressions makes their numerical implementation still extremely difficult.

%
In this work, the Coulomb gyrokinetic collision operator is derived in a form that can be efficiently implemented in numerical simulation codes as it involves only integrals over the two gyrokinetic velocity coordinates.
The derivation of the full Coulomb collision operator is based on a multipole expansion of the Rosenbluth potentials.
This allows us to write the Coulomb collision operator in terms of moments of the distribution function and apply the gyroaverage operator to the resulting expansion.
The Coulomb collision operator {is} then expressed in terms of two-dimensional velocity integrals of the  distribution function.
We show that the gyroangle dependence of the expansion coefficients, given in terms of scalar spherical harmonics, allows for analytical gyroaveraging integrations at arbitrary values of the perpendicular wavevector.
Furthermore, motivated by recent work based on a pseudo-spectral approach to the gyrokinetic equation \citep{Mandell2018,Frei2019}, the collision operator is projected onto a Hermite-Laguerre polynomial basis, and is expressed in terms of moments of the distribution function on the same basis.
{The set of moment-hierarchy equations can then be rigorously closed by using systematic techniques [such as the semi-collisional closure \citep{Zocco2011,Jorge2017}] without requiring ad-hoc truncation of infinite series.}

This paper is organized as follows. 
\cref{sec:ordering} derives the gyrokinetic equation and \cref{sec:gcmodel} presents the multipole expansion of the Coulomb collision operator.
In \cref{sec:gudingcentertransf}, the Coulomb operator is ported to a gyrocenter coordinate system, while \cref{eq:hermlag} projects the collision operator onto a Hermite-Laguerre polynomial basis to obtain a closed-form expression of the Coulomb collision operator in terms of the moments of the distribution function on the same basis.
The gyrokinetic collision operator for unlike-species is presented in \cref{sec:smallmassratio} using an expansion based on the smallness of the electron-to-ion mass ratio.
The conclusions follow.

\section{Gyrokinetic Model}
\label{sec:ordering}

The evolution of the distribution function $f_a=f_a(\mathbf x, \mathbf v)$ is given by the Boltzmann equation
\begin{equation}
    \frac{\partial f_a}{\partial t}+\dot{\mathbf x}\cdot \frac{\partial f_a}{\partial \mathbf x}+\dot{\mathbf v} \cdot \frac{\partial f_a}{\partial \mathbf v} = \sum_b C(f_a,f_b),
\label{eq:boltzmanne}
\end{equation}
with $C(f_a,f_b)$ the Coulomb (also known as Landau) collision operator \citep{Landau1936,Rosenbluth1957}.
This is an operator of the Fokker-Planck type, derived from first principles, and valid in the common case where small-angle Coulomb collisions dominate.
Its expression is given by
\begin{equation}
    \begin{split}
        C(f_a,f_b)&=L_{ab} \sum_{j,k} \frac{\partial}{\partial v_k}\l[\frac{\partial}{\partial v_j}\left(\frac{\partial^2 G_b}{\partial v_k \partial v_j}f_a\right)-2\left(1+\frac{m_a}{m_b}\right)\frac{\partial H_b}{\partial v_k}f_a\r],
    \end{split}
    \label{eq:coulombop}
\end{equation}
with
\be
    \begin{split}
        H_b&=\int \frac{f_b(\mathbf v')}{|\mathbf v - \mathbf v'|}d\mathbf v',
    \end{split}
\label{eq:roshb}
\ee
and
\be
    \begin{split}
        G_b=\int f_b(\mathbf v')|\mathbf v - \mathbf v'|d\mathbf v',
    \end{split}
\ee
the Rosenbluth potentials satisfying $\nabla^2_v G_b =2 H_b$ and $\nabla^2_v H_b = - 4 \pi f_b$ \citep{Rosenbluth1957}. 
In \cref{eq:coulombop}, $L_{ab}=q_a^2 q_b^2 \lambda_{ab}/(8 \pi \epsilon_0^2 m_a^2)=\nu_{ab} v_{tha}^3/n_b$ is introduced, where $\lambda_{ab}$ and $\nu_{ab}$ are the Coulomb logarithm and the collision frequency between species $a$ and $b$ respectively, $v_{tha}=\sqrt{2 T_a/m_a}$ is the thermal {speed}, and $q_a$ and $m_a$ are the charge and the mass of particles of species $a$, $a=e,i$.

In the present paper we consider a plasma with properties that satisfy the gyrokinetic ordering \citep{Brizard2009,Frei2019}.
More precisely, denoting typical turbulent frequencies as $\omega \sim |\partial_t \log n|\sim |\partial_t \log T_e|$ with $n_e$ and $T_e$ the electron density and temperature respectively, and typical wavenumbers $\mathbf k = k_\parallel \mathbf b + \mathbf k_\perp$, being $\mathbf k \sim |\nabla \log n_e|\sim |\nabla \log T_e|$ and $\mathbf b = \mathbf B/B$ the magnetic field unit vector, we assume
\begin{equation}
    \epsilon \sim \frac{|\mathbf v_{\mathbf E}|}{c_s} \sim \frac{k_\parallel}{k_\perp} \ll 1,
\label{eq:ordering}
\end{equation}
where $c_s=\sqrt{T_e/m_i}$ is the sound speed, $\rho_s=c_s/\Omega_i$ the sound Larmor radius, $\Omega_i=eB/m_i$ the ion gyrofrequency, and $\mathbf v_{\mathbf E}=\mathbf E\times \mathbf B/B^2$ the $\mathbf E \times \mathbf B$ drift velocity with $\mathbf E = - \nabla \phi - \partial_t \mathbf A$ the electric field.
As strong radial electric fields are known to play a role in the tokamak edge (particularly in the H-mode pedestal), and large scale fluctuations are the ones at play in the tokamak scrape-off layer, we split the electrostatic potential as $\phi=\phi_0+\phi_1$ \citep{Dimits1992,Qin2006,Frei2019}, i.e. into a possibly large-scale drift-kinetic component, $\phi_0$, satisfying
\begin{equation}
    \frac{e \phi_0}{T_e} \sim 1,
\label{eq:ordering2}
\end{equation}
and its small-amplitude gyrokinetic, $\phi_1$, component
\begin{equation}
    \frac{\phi_1}{\phi_0} \sim \epsilon_\delta \ll 1.
\label{eq:phi10}
\end{equation}
A similar decomposition into large and small scale fluctuations is applied to the magnetic vector potential $\mathbf A=\mathbf A_0+\mathbf A_1$, with $|\mathbf A_1|/|\mathbf A_0|\sim \epsilon_\delta$.
Both $\phi_0$ and $\phi_1$ are assumed to yield a similar contribution to the total electric field
\begin{equation}
    \mathbf E \sim \nabla_\perp \phi_0 \sim \nabla_\perp \phi_1.
\label{eq:ephi10}
\end{equation}
Therefore, by ordering typical gradient lengths of $\phi_1$ to be comparable to $\rho_s$, we set
\begin{equation}
    \rho_s \left|\frac{\nabla_\perp \phi_1}{\phi_1}\right| \sim 1,
\label{eq:kperprhos1}
\end{equation}
which, using \cref{eq:phi10,eq:ephi10}, constraints typical gradient lengths of $\phi_0$ to be much larger than $\rho_s$, as
\begin{equation}
    \rho_s \left|\frac{\nabla_\perp \phi_0}{\phi_0}\right| \sim \epsilon_\delta.
\label{eq:kperprhos2}
\end{equation}
In the following, we set $\epsilon_\delta \sim \epsilon$, {which, using \cref{eq:ordering}, yields}
\begin{equation}
	{k_\perp \rho_s \frac{e \phi_0}{T_e}\sim \frac{e \phi_1}{T_e} \sim \epsilon.}
\end{equation}
The scale length $L_B \sim R_0$ of the equilibrium magnetic field (with $R_0$ the major radius of the tokamak device), is ordered by the small parameter $\epsilon_B\sim{\rho_s}/{L_B}$.
{We note that the collision operator developed here is valid for both $\epsilon_B \sim \epsilon^2$ and $\epsilon_B \sim \epsilon^3$, with the second case being more of interest for the periphery since the plasma temperature is lower than the tokamak core \citep{Hahm2009}.}
Finally, the collision frequency is ordered as
\begin{equation}
    \frac{\nu_i}{\Omega_i}\sim \epsilon_\nu \sim \epsilon^2,
\label{eq:orderingcoll}
\end{equation}
with $\nu_i=\nu_{ii}$ the ion-ion collision frequency.
For $T_e \sim T_i$, the ordering in \cref{eq:orderingcoll} implies that $k_\parallel \lambda_{mfpe} \sim k_\parallel \lambda_{mfpi} \sim k_\perp \rho_s/\epsilon$, with $\lambda_{mfpa} = v_{tha}/\nu_a$.

By taking advantage of the ordering in Eqs. (\ref{eq:ordering}-\ref{eq:phi10}) and \cref{eq:orderingcoll}, the gyrokinetic model effectively removes the fast time scale associated with the cyclotron motion and reduces the dimensionality of the kinetic equation from six phase-space variables, i.e. ($\mathbf x, \mathbf v$), to five.
While linear and nonlinear gyrokinetic equations of motion were originally derived using recursive techniques \citep{Taylor1968,Rutherford1968a,Catto1978a}, more recent derivations of the gyrokinetic equation based on Hamiltonian Lie perturbation theory \citep{Cary1981} ensure the existence of phase-space volume and magnetic moment conservation laws \citep{Hahm1988a,Brizard2007a,Hahm2009,Frei2019}.
The Hamiltonian derivations are carried out in two steps. In the first step, small-scale electromagnetic fluctuations with perpendicular wavelengths comparable to the particle Larmor radius ($\phi_1$ and $A_{\parallel 1}$) are neglected \citep{Cary2009}.
Within this approximation, the coordinate transformation from particle phase-space coordinates ($\mathbf x, \mathbf v)$ to guiding-center coordinates $\mathbf Z = (\mathbf R, v_\parallel, \mu, \theta)$ is derived, where $\mathbf R$ is the guiding-center, $v_\parallel$ the parallel velocity, $\mu$ the {magnetic moment}, and $\theta$ the gyroangle.
The second step introduces small-scale and small-amplitude electromagnetic fluctuations, $\phi_1$ and $\mathbf A_1$.
For this purpose, a gyrocenter coordinate system $\overline{\mathbf Z}=(\overline{\mathbf R},\overline{v}_\parallel, \overline{\mu}, \overline{\theta})$ is constructed perturbatively from the guiding-center coordinates $\mathbf Z$ via a transformation $T$ of the form
\begin{equation}
    \overline{\mathbf Z} = T \mathbf Z = \mathbf Z + \epsilon_\delta \mathbf Z_1 + ...,
\label{eq:gyrotransf}
\end{equation}
where $\mathbf Z_1$ contains terms proportional to $\phi_1$ and $\mathbf A_1$, such that $\overline \mu = T \mu = \mu + \epsilon_\delta \mu_1 + ...$, remains an adiabatic invariant [see, e.g., \citet{Brizard2007a}].
This allows us to reduce the number of phase-space variables in the kinetic Boltzmann equation describing the evolution of the particle distribution function from six to five, simplifying the analytical and numerical treatment of magnetized plasma systems.

More precisely, in order to simplify and remove the gyroangle dependence of the Boltzmann equation, the distribution function $f_a(\mathbf x, \mathbf v)$ is first expressed in terms of the guiding-center coordinates $\mathbf Z$ by defining the guiding centre distribution function $F_a(\mathbf Z)$ as
\begin{equation}
    F_a(\mathbf Z) = f_a(\mathbf x(\mathbf Z), \mathbf v(\mathbf Z)).
\label{eq:gdpf}
\end{equation}
The coordinate transformation $\mathbf v(\mathbf Z)$ is given by
\begin{equation}
    \mathbf v = v_\parallel \mathbf b { + \mathbf v_{E{0}}}+v_\perp{'}(\cos \theta \mathbf e_1 + \sin \theta \mathbf e_2)=v\left[\cos \varphi \mathbf b + \sin \varphi(\cos \theta \mathbf e_1 + \sin \theta \mathbf e_2)\right],
\label{eq:gcv}
\end{equation}
with $(\mathbf b, \mathbf e_1, \mathbf e_2)$ a fixed right-handed coordinate set, {$v_{E0}=-\nabla \phi_0$ the drift-kinetic $\mathbf E\times \mathbf B$ drift,} $\cos \varphi=v_\parallel/v$ the cosine of the pitch angle and $\theta$ the gyroangle.
The {magnetic} moment $\mu$ is defined as
\begin{equation}
    \mu = \frac{m_a v_\perp^{{'}2}}{2 B},
\label{eq:gcmu}
\end{equation}
whereas the particle position $\mathbf x(\mathbf Z)$ is written as
\begin{equation}
    \mathbf x = \mathbf R + \mathbf \rho_a,
\label{eq:gcx}
\end{equation}
with
\begin{equation}
    \mathbf \rho_a = \mathbf \rho_a(\mathbf R, \mu, \theta) = \sqrt{\frac{2m_a\mu}{q_a^2 B}}(-\sin \theta \mathbf e_1 + \cos \theta \mathbf e_2)
\end{equation}
the Larmor radius and $\mathbf R$ the guiding-center of the particle.
The Jacobian of the guiding-center transformation of Eqs. (\ref{eq:gcv}-\ref{eq:gcx}) is given by $B_\parallel^*/m_a = (B/m_a)(1 {+\mathbf b \cdot \nabla \times \mathbf v_E/\Omega_a}+v_\parallel \mathbf b \cdot \nabla \times \mathbf b/\Omega_a)$ {\citep{Cary2009}.}
{
We note that, in a weak-flow regime where $\mathbf v_E$ is absent from \cref{eq:gcv}, the calculation that follows remains valid, except for the Jacobian of the guiding-center transformation $B_{\parallel}^*$, which should be replaced by $B_{\parallel}^*= (B/m_a)(1+v_\parallel \mathbf b \cdot \nabla \times \mathbf b/\Omega_a)$.
}

To account for the small scale fluctuations {and magnetic inhomogeneity}, the gyrokinetic distribution function $\overline{F}_a(\overline{\mathbf{Z}})$ is then defined as \citep{Brizard2007a}
\begin{equation}
    \overline{F_a}(\overline{\mathbf Z}) = F_a(\mathbf Z){=T \overline F_a(\mathbf Z)},
\end{equation}
with the coordinate transformation between $\overline{\mathbf Z}$ and ${\mathbf Z}$ given perturbatively by \cref{eq:gyrotransf}.
Indeed, using the chain rule to rewrite the Boltzmann equation, \cref{eq:boltzmanne}, in terms of gyrocenter $\overline{\mathbf Z}$ coordinates, we obtain
\begin{equation}
    \frac{\partial \overline{F_a}}{\partial t} + \dot{\overline {\mathbf Z}} \cdot \frac{\partial \overline{F_a}}{\partial \overline {\mathbf Z}}= \sum_b C(\overline{F_a},\overline{F_b}).
\label{eq:gykboltz}
\end{equation}
We now introduce the gyroaverage operator $\left< ... \right>_{\overline{\mathbf R}}$ defined by
\begin{equation}
    \left< \chi \right>_{\overline{\mathbf R}} = \frac{1}{2\pi}\int_0^{2\pi}\chi(\overline{\mathbf Z})d\overline \theta,
\label{eq:gkgyro}
\end{equation}
where all gyrocenter coordinates $\overline{\mathbf Z}$ but $\overline \theta$ are kept fixed during the integration.
By applying the gyroaverage operator to \cref{eq:gykboltz}, the gyrokinetic equation is obtained 
\begin{equation}
    \frac{\partial}{\partial t}\left< \overline{F_a} \right>_{\overline{\mathbf R}} + \left<\dot{\overline {\mathbf Z}} \cdot \frac{\partial \overline{F_a}}{\partial \overline {\mathbf Z}} \right>_{\overline{\mathbf R}}=\sum_b \left< C(\overline{F_a},\overline{F_b})\right>_{\overline{\mathbf R}}.
\label{eq:boltzmann1}
\end{equation}
Equation (\ref{eq:boltzmann1}) can be further simplified by noting that, in the gyrokinetic framework, the transformation in \cref{eq:gyrotransf} is constructed in such a way that the gyrocenter equations of motion, i.e. the equations that determine $\dot{\overline {\mathbf Z}}=(\dot{\overline{\mathbf R}},\dot{\overline{v_\parallel}},\dot{\overline{\mu}},\dot{\overline{\theta}})$, are gyroangle independent and that $\overline \mu$ is an adiabatic invariant satisfying $\dot{\overline \mu}=0$ \citep{Brizard2007a}.
Therefore, the gyrokinetic equation in \cref{eq:boltzmann1} can be written as
\begin{equation}
    \frac{\partial}{\partial t}\left< \overline{F_a} \right>_{\overline{\mathbf R}} + \dot{\overline {\mathbf R}} \cdot \frac{\partial \left<\overline{F_a}\right>_{\overline{\mathbf R}}}{\partial \overline {\mathbf R}}+ \dot{\overline {v_\parallel}} \frac{\partial \left<\overline{F_a}\right>_{\overline{\mathbf R}}}{\partial \overline {v_\parallel}} = \sum_b\left< C(\overline{F_a},\overline{F_b})\right>_{\overline{\mathbf R}}.
\label{eq:boltzmann}
\end{equation}

In order to further simplify \cref{eq:boltzmann}, we estimate the order of magnitude of the gyrophase dependent part of the distribution function $\widetilde{\overline{F_a}} =\overline{F_a} - \left< \overline{F_a} \right>_{\overline{\mathbf{R}}}$, where $\overline{F_a}$ obeys \cref{eq:gykboltz} and $\left< \overline{F_a} \right>_{\overline{\mathbf{R}}}$ obeys \cref{eq:boltzmann}.
For this purpose, we note that the equation for the evolution of $\widetilde{\overline{ F_a}}=\overline{F_a}-\left< \overline{F_a} \right>_{\overline{\mathbf{R}}}$ can be obtained by subtracting \cref{eq:boltzmann} from \cref{eq:gykboltz}, yielding
\begin{equation}
    \frac{\partial \widetilde{\overline{ F_a}}}{\partial t} + \dot{\overline {\mathbf R}} \cdot \frac{\partial \widetilde{\overline{ F_a}}}{\partial \overline {\mathbf R}} +\dot{\overline {v_\parallel}} \frac{\partial \widetilde{\overline{ F_a}}}{\partial \overline {v_\parallel}}+\dot{\overline {\theta}} \frac{\partial \widetilde{\overline{ F_a}}}{\partial \overline {\theta}} = \sum_b C(\overline{F_a},\overline{F_b})-\left<C(\overline{F_a},\overline{F_b})\right>_{\overline{\mathbf{R}}}.
\label{eq:gykboltz1}
\end{equation}
To lowest order, $\dot{\overline {\theta}} \partial_\theta \widetilde{\overline{F_a}} \sim \Omega_a \widetilde{\overline{F_a}} $ and $\partial_t \sim \dot{\overline {\mathbf R}} \cdot \nabla_{ \overline {\mathbf R}}\sim \dot{\overline{v_\parallel}}\partial_{{\overline{v_\parallel}}}\sim \epsilon \Omega_i$.
Therefore, the leading order estimate of \cref{eq:gykboltz1} gives
\begin{equation}
    \widetilde {\overline{F_a}} \simeq \frac{1}{\Omega_a} \sum_b \int_0^{\overline \theta} \left[C\left(\left<\overline{F_a}\right>_{\overline{\mathbf{R}}},\left<\overline{F_b}\right>_{\overline{\mathbf{R}}}\right)-\left<C\left(\left<\overline{F_a}\right>_{\overline{\mathbf{R}}},\left<\overline{F_b}\right>_{\overline{\mathbf{R}}}\right)\right>_{\overline{\mathbf{R}}}\right]d{\overline{\theta}'}.
\label{eq:leadingorderftilde}
\end{equation}
Using the fact that $C(\overline{F_a},\overline{F_b}) \sim \nu_a \overline F_a$, together with \cref{eq:orderingcoll}, and expanding $\overline {F_a}$ as $\overline {F_a}=\left< \overline{ F_a} \right>_{\overline{\mathbf{R}}}+\epsilon_\nu \overline {F_{a1}}+...$, we find that
\begin{equation}
    \frac{\widetilde {\overline{F_e}}}{\left< \overline{ F_e} \right>_{\overline{\mathbf{R}}}} \sim \left(\frac{T_i}{T_e}\right)^{3/2}\sqrt{\frac{m_e}{m_i}}\epsilon_\nu \sim \left(\frac{T_i}{T_e}\right)^{3/2} \sqrt{\frac{m_e}{m_i}}\epsilon^2,
    \label{eq:orderingftildee}
\end{equation}
and
\begin{equation}
    \frac{\widetilde {\overline F_i}}{\left< \overline F_i \right>_{\overline{\mathbf{R}}}} \sim \frac{\nu_i}{\Omega_i}\sim \epsilon_\nu \sim \epsilon^2.
    \label{eq:orderingftildei}
\end{equation}
showing that, up to second order in $\epsilon$, the gyroangle dependence of the distribution function can be neglected in \cref{eq:boltzmann}.
We remark that a similar estimate for the gyrophase dependent part of the guiding-center distribution function was found in \citet{Jorge2017}.

We now evaluate the magnitude of the collisional term in \cref{eq:gykboltz1}.
Using the expansion $C(\overline{F_a},\overline{F_b}) = C_0(\overline{F_a},\overline{F_b}) + \epsilon_\delta C_1(\overline{F_a},\overline{F_b})+...$ with $C_0(\overline{F_a},\overline{F_b}) \sim \nu_a \overline{F_a}$, and noting that the first order gyrocenter transformation $\mathbf Z_1=\mathbf Z - \overline{\mathbf Z}+O(\epsilon_\delta^2)$ in \cref{eq:gyrotransf} is mass dependent, i.e. $\overline{\mathbf Z_{1e}} \sim \sqrt{m_e/m_i} \overline{\mathbf Z_{1i}}$ [see, e.g., \citet{Brizard2007a}], the magnitude of the Coulomb collision operator for electrons can be estimated as
\begin{equation}
    C(\overline F_e,\overline{F_b}) \sim \nu_e \overline F_e \sim \sqrt{\frac{m_i}{m_e}} \epsilon_\nu \Omega_i \overline F_e+O\left(\epsilon_\nu \epsilon_\delta \Omega_i \overline F_e\right)\sim \sqrt{\frac{m_i}{m_e}} \epsilon^2 \Omega_i \overline F_e+O\left(\epsilon^3\Omega_i \overline F_e\right).
\label{eq:orderingcfe}
\end{equation}
{Thus, the third order term in the expansion in \cref{eq:orderingcfe} does not contain a $\sqrt{m_i/m_e}$ factor, in contrast with its lower order counterpart.}
A similar argument holds for the ions, yielding
\begin{equation}
    C(\overline F_i,\overline{F_b}) \sim \nu_i \overline F_i \sim \epsilon_\nu \Omega_i \overline F_i +O\left(\epsilon_\nu \epsilon_\delta \Omega_i \overline F_i\right)\sim \epsilon^2 \Omega_i \overline F_i +O\left(\epsilon^3 \Omega_i \overline F_i\right).
\label{eq:orderingcfi}
\end{equation}
Equations (\ref{eq:orderingcfe}) and (\ref{eq:orderingcfi}) show that the lowest order collision operator $C_0(\overline{F_a},\overline{F_b})$ is, in fact, $O(\epsilon^2)$.
%
Therefore, the gyrokinetic equation valid up to second order in $\epsilon$, considered in a large number of edge gyrokinetic models [see, e.g., \citet{Qin2006,Qin2007,Hahm2009,Frei2019}], can thus be written as
\begin{equation}
    \frac{\partial}{\partial t}\left< \overline{F_a} \right>_{\overline{\mathbf{R}}} + \dot{\overline {\mathbf R}} \cdot \frac{\partial }{\partial \overline {\mathbf R}}\left<\overline{F_a}\right>_{\overline{\mathbf{R}}}+ \dot{\overline {v_\parallel}} \frac{\partial}{\partial \overline {v_\parallel}}\left<\overline{F_a}\right>_{\overline{\mathbf{R}}} = \sum_b \left< C_0(\left<\overline{F_a}\right>_{\overline{\mathbf{R}}},\left<\overline{F_b}\right>_{\overline{\mathbf{R}}})\right>_{\overline{\mathbf{R}}},
\label{eq:boltzmannfinal}
\end{equation}
assuming $\dot{\overline {\mathbf R}}$ and $\dot{\overline {v_\parallel}}$ to be at least $O(\epsilon^2)$ accurate.
We note that although only the lowest order in $\epsilon_\delta$ collision operator $C_0(\left<\overline{F_a}\right>_{\overline{\mathbf{R}}},\left<\overline{F_b}\right>_{\overline{\mathbf{R}}})$ is used in \cref{eq:boltzmannfinal}, all orders in $k_\perp \rho_s$ are kept.

\section{Multipole Expansion of the Coulomb Collision Operator}
\label{sec:gcmodel}

The goal of this section is to find a suitable basis to expand $f_a$ such that the Coulomb operator in \cref{eq:coulombop} can be cast as a function of moments of $f_a$.
This first step considerably simplifies the derivation of the gyrokinetic collision operator.
We start by noting that the Rosenbluth potential $H_b$ in \cref{eq:roshb} is analogous to the expression of the electrostatic potential due to a charge distribution, a similarity already noted by \citet{Rosenbluth1957}.
This fact allows us to make use of known electrostatic expansion techniques \citep{Jackson1999} to perform a multipole expansion of the Rosenbluth potentials.
We first Taylor expand the factor $1/|\mathbf v - \mathbf v'|$ in \cref{eq:roshb} around $\mathbf v=0$ if $v\le v'$ or around $\mathbf v'=0$ if $v>v'$, yielding
\begin{equation}
\frac{1}{|\mathbf v - \mathbf v'|} =
\begin{cases}
    \mathlarger{\sum}\limits_{l=0}^\infty \dfrac{(-\mathbf v')^l}{l!} \cdot \dfrac{\partial^l}{\partial {\mathbf v}^l} \left(\dfrac{1}{v}\right), &v'\le v,\\
    \mathlarger{\sum}\limits_{l=0}^\infty \dfrac{(-\mathbf v)^l}{l!} \cdot \dfrac{\partial^l}{\partial {(\mathbf v')}^l} \left(\dfrac{1}{v'}\right), &v<v'.\\
\end{cases}
\end{equation}
where we used the identity $\partial_{\mathbf v}(1/|\mathbf v - \mathbf v'|)_{v=0}=-\partial_{\mathbf v'}(1/v')$ and where we denote the inner product between all the $l$ indices of two $l$-rank tensors, $\mathbf T_1^l$ and $\mathbf T_2^l$, as $\mathbf T_1^l \cdot \mathbf T_2^l$.
Both $v\le v'$ and $v>v'$ cases are included in order to take into account the fact that $f_b(\mathbf v')$ is, in general, finite over the entire velocity space $\mathbf v'$.
Denoting $\mathbf Y^{l}(\mathbf v)$ the spherical harmonic tensor \citep{Weinert1980}
\begin{equation}
    \mathbf Y^{l}(\mathbf v) = \frac{(-1)^l v^{2l+1}}{(2l-1)!!}\left(\frac{\partial}{\partial \mathbf v}\right)^l \frac{1}{v},
\label{eq:yltensor}
\end{equation}
we obtain the following expression for $H_b$
\begin{equation}
    H_b = 2 \sum_{l=0}^\infty \frac{(2l-1)!!}{l!} \left(\int_{v>v'} f_b(\mathbf v') \frac{(\mathbf v')^l}{v^{2l+1}} \cdot \mathbf Y^{l}(\mathbf v)  d\mathbf v'+\int_{v'\ge v} f_b(\mathbf v') \frac{(\mathbf v)^l}{(v')^{2l+1}} \cdot \mathbf Y^{l}(\mathbf v')  d\mathbf v'\right).
\label{eq:roshb1}
\end{equation}
%
%

In order to simplify \cref{eq:roshb1}, we note that the tensor $\mathbf Y^{l}(\mathbf v)=Y_{\alpha \beta ... \gamma}^l(\mathbf v)$ is symmetric and totally traceless, i.e. traceless between any combination of two of its indices.
Symmetry arises from the fact that any couple of indices in $Y_{\alpha \beta ... \gamma}^l(\mathbf v)$ is interchangeable as the velocity derivatives commute for $v \not= 0$.
The totally traceless feature (i.e. $\sum_\alpha Y_{\alpha \alpha ... \gamma}^l(\mathbf v)=0$ and the same for any other pair of indices), stems from the fact that the contraction between any two indices in $Y_{\alpha \beta ... \gamma}^l(\mathbf v)$ leads to the multiplicative factor $\nabla_{\mathbf v}^2=\partial_{\mathbf v}\cdot \partial_{\mathbf v} (1/v)$, which vanishes for $v\not=0$ {(we note that $Y^l(\mathbf v)$ vanishes in the limit $\mathbf v=0$).}
Furthermore, by defining the tensor $(\mathbf v)^l_{\text{TS}}$ as the traceless symmetric counterpart of $(\mathbf v)^l$ [e.g., $(\mathbf v)^2_{\text{TS}}=\mathbf v \mathbf v - \mathbf I v^2/3$ with $\mathbf I$ the identity matrix], we replace the tensors $(\mathbf v')^l$ and $(\mathbf v)^l$ in \cref{eq:roshb1} by their traceless symmetric counterpart $(\mathbf v')^l_{\text{TS}}$ and $(\mathbf v)^l_{\text{TS}}$ respectively
\begin{equation}
    H_b = 2 \sum_{l=0}^\infty \frac{(2l-1)!!}{l!} \left(\int_{v>v'} f_b(\mathbf v') \frac{(\mathbf v')^l_{\text{TS}}}{v^{2l+1}} \cdot \mathbf Y^{l}(\mathbf v)  d\mathbf v'+\int_{v'\ge v} f_b(\mathbf v') \frac{(\mathbf v)^l_{\text{TS}}}{(v')^{2l+1}} \cdot \mathbf Y^{l}(\mathbf v')  d\mathbf v'\right),
\label{eq:roshb11}
\end{equation}
as they differ only by terms proportional to the identity matrix that vanish when summed with $\mathbf Y^{l}(\mathbf v)$ and $\mathbf Y^{l}(\mathbf v')$.
In fact, for the $l=2$ case we have $(\mathbf v^2-(\mathbf v)^2_{\text{TS}})\cdot \mathbf Y^2(\mathbf v)=(v^2/3)\mathbf I \cdot \mathbf Y^2(\mathbf v)=(v^2/3) \sum_\alpha Y^2_{\alpha \alpha} =0$, and similarly for $l>2$.
In addition, following the convention in \citet{Snider2018}, scalars ($l=0$) and vectors ($l=1$) are considered to be traceless symmetric quantities.
Finally, we relate the tensors $(\mathbf v)^l_{\text{TS}}$ and $\mathbf Y^{l}(\mathbf v)$.
For $l=0$ and $l=1$, we have $\mathbf Y^{0}(\mathbf v')=1=(\mathbf v')^0_{\text{TS}}$ and $\mathbf Y^{1}(\mathbf v')=\mathbf v'=(\mathbf v')^1_{\text{TS}}$.
For $l=2$, \cref{eq:yltensor} gives
\begin{equation}
\begin{split}
    \mathbf Y^2(\mathbf v') &= \mathbf v' \mathbf v' - \frac{v'^2}{3}\mathbf I=(\mathbf v')^2_{TS}.
\end{split}
\label{eq:yl2}
\end{equation}
%
%
The results obtained for $l=0,1$, and 2 can be generalized, i.e. $(\mathbf v')^l_{\text{TS}}=\mathbf Y^{l}(\mathbf v')$ as proved by induction \citep{Weinert1980}.
The Rosenbluth potential $H_b$ can therefore be written as
\begin{equation}
    H_b = 2 \sum_{l=0}^\infty \frac{(2l-1)!!}{l!}\mathbf Y^{l}(\mathbf v) \cdot \left[\frac{1}{(v^{2})^{l+1/2}} \int_{v'<v} f_b(\mathbf v') \mathbf Y^{l}(\mathbf v') d\mathbf v'+ \int_{v'\ge v} f_b(\mathbf v') \frac{\mathbf Y^{l}(\mathbf v')}{[(v')^{2}]^{l+1/2}}  d\mathbf v'\right].
\label{eq:roshb2}
\end{equation}
The first term in \cref{eq:roshb2} can be regarded as the potential due to the charge distribution $f_b(\mathbf v')$ inside a sphere of radius $v$, while the second term is the potential due to a finite charge distribution $f_b(\mathbf v')$ at $v'\ge v$.

We now look for an expansion of $f_b$ that allows us to perform the integrals in \cref{eq:roshb2} analytically by writing $H_b$ as a sum of velocity moments of $f_b$.
We consider the basis functions \citep{Hirshman1976a}
\begin{equation}
    \mathbf Y^{lk}(\mathbf v) = \mathbf Y^{l}\left(\mathbf v\right) L_k^{l+1/2}(v^2),
\end{equation}
with $L_k^{l+1/2}(v)$ an associated Laguerre polynomial \citep{Abramowitz1972}, i.e.
\be
    \begin{split}
        L_{k}^{l+1/2}(v)&=\sum_{m=0}^{k}L_{km}^{l} v^m,
    \end{split}
    \label{eq:asslaguerre}
\ee
where
\begin{equation}
    L_{km}^{l}=\frac{(-1)^m(l+k+1/2)!}{(k-m)!(l+m+1/2)!m!}.
\end{equation}
The basis $\mathbf Y^{lk}(\mathbf v)$ is orthogonal, being the orthogonality relation given by \citep{Banach1989,Snider2018}
\begin{equation}
    \int e^{-v^2} \mathbf Y^{l'k'}(\mathbf v) \mathbf Y^{lk}(\mathbf v) d \mathbf v \cdot {\mathbf T^{lk}}= \delta_{ll'}\delta_{kk'} \pi^{3/2} \sigma_{k}^l {\mathbf T^{lk}},
\label{eq:orthoy}
\end{equation}
with $\mathbf T^{lk}$ an arbitrary symmetric and traceless tensor, and $\sigma_k^l$ the normalization constant
\begin{equation}
    \sigma_k^l=\frac{l!(l+k+1/2)!}{2^l(l+1/2)!k!}.
\end{equation}
A proof that $\mathbf Y^{lk}(\mathbf v)$ is a complete basis, i.e. that each $l$ and $k$ element of $\mathbf Y^{lk}(\mathbf v)$ is linearly independent and that a linear combination of its elements spans any smooth function $f(\mathbf v)$, can be found in \citet{Banach1989}, where the equivalence between Grad's moment expansion in tensorial Hermite polynomials (which forms a complete basis) and $\mathbf Y^{lk}(\mathbf v)$ is shown.
We then write $f_b$ as
\begin{equation}
    f_b = f_{Mb}\sum_{l,k=0}^{\infty} \mathbf Y^{lk}\left(\frac{\mathbf v}{v_{thb}}\right) \cdot \frac{\mathbf M_b^{lk}}{\sigma_{k}^l},
\label{eq:faexp}
\end{equation}
with $f_{Mb}$ a Maxwellian distribution function
\begin{equation}
    f_{Mb}=\frac{n_b}{v_{thb}^3 \pi^{3/2}} e^{-\frac{v^2}{v_{thb}^2}}.
\end{equation}
According to \cref{eq:orthoy}, with ${\mathbf T^{lk}}=\mathbf M^{lk}_b$, the coefficients $\mathbf M^{lk}_b$ are obtained by taking velocity moments of $f_b$ of the form
\begin{equation}
    \mathbf M^{lk}_b=\frac{1}{n_b}\int f_b(\mathbf v) \mathbf Y^{lk}\left(\frac{\mathbf v}{v_{thb}}\right) d\mathbf v.
\label{eq:defmaln}
\end{equation}
Finally, we note that \cref{eq:faexp} allows us to retain only the $l=k=0$ moment when the plasma is in thermal equilibrium.

Plugging the expansion for $f_b$ given by \cref{eq:faexp} into \cref{eq:roshb2}, we obtain the following expression
\begin{align}
    H_b &= \frac{n_b}{v_{thb} \pi^{3/2}} \sum_{l,l',k} \frac{(2l-1)!!}{l!\sigma_k^{l'}}\nonumber\\
    &\times\left(\frac{\mathbf Y^{l}(\hat v)}{x_b^{(l+1)/2}} \cdot \int_{0}^{x_b} e^{-x} L_k^{l'+1/2}(x) x^{(l+l'+1)/2} dx \int \mathbf Y^{l}(\hat v')\mathbf Y^{l'}(\hat v') d\sigma' \cdot \mathbf M_b^{l'k}\right.\nonumber\\
    &\left.+x_b^{l/2} \mathbf Y^{l}(\hat v) \cdot \int_{x_b}^{\infty} e^{-x}L_k^{l+1/2}(x) dx \int \mathbf Y^{l}(\hat v')\mathbf Y^{l'}(\hat v') d\sigma' \cdot \mathbf M_b^{l'k}\right),
\label{eq:roshb3}
\end{align}
where we define the normalized velocity $x_b=v^2/v_{thb}^2$, the solid angle $\sigma$ such that $d \mathbf v = v^2 dv d\sigma$, and use the relation $\mathbf Y^{l} \mathbf (v) = v^l \mathbf Y^l(\hat v)$ with $\mathbf v = v \hat v$ \citep{Weinert1980}.
Applying the orthogonality relation of \cref{eq:orthoy} for $k=0$, and expanding the associated Laguerre polynomials using \cref{eq:asslaguerre}, we write $H_b$ as
\begin{align}
    H_b &= \frac{2n_b}{v_{thb}} \sum_{l,k} \sum_{m=0}^k \frac{L_{km}^{l}}{\sigma_k^l}\frac{\mathbf Y^{l}(\hat v) \cdot \mathbf M_b^{lk}}{2l+1} \nonumber\\
    &\times\frac{1}{\sqrt{\pi}}\left(\frac{1}{x_b^{(l+1)/2}} \int_{0}^{x_b} e^{-x} x^{m+l+1/2} dx+x_b^{l/2} \int_{x_b}^{\infty} e^{-x}x^m dx\right),
\label{eq:roshb4}
\end{align}
where the identity
\begin{equation}
    \frac{(2l-1)!!}{2^l (l+1/2)!}=\frac{2}{\sqrt{\pi}}\frac{1}{2l+1},
\end{equation}
is used to simplify \cref{eq:roshb4}.

We note that the expression of $H_b$ in \cref{eq:roshb4} corresponds to the one in \citet{Ji2006}, having replaced the $\mathbf Y^l(\mathbf v)$ tensors by the $\mathbf P^l(\mathbf v)$ tensors defined by the recursion relation [see Eq. (14) of \citet{Ji2006}]
\begin{equation}
    \mathbf P^{l+1}(\mathbf v) = \mathbf v \mathbf P^{l}(\mathbf v) -\frac{v^2}{2l+1}\frac{\partial}{\partial \mathbf v}\mathbf P^{l}(\mathbf v),
\label{eq:recpl}
\end{equation}
with $\mathbf P^0(\mathbf v)=1$ and $\mathbf P^1(\mathbf v)=\mathbf v$.
We can indeed prove that $\mathbf Y^l(\mathbf v) = \mathbf P^l (\mathbf v)$ by deriving the tensor $\mathbf Y^{l}(\mathbf v)$ using \cref{eq:yltensor}.
This yields
\begin{align}
    \frac{\partial}{\partial \mathbf v}\mathbf Y^l(\mathbf v) &= \frac{(-1)^l}{(2l-1)!!}\left[(2l+1)v^{2l-1}\mathbf v \frac{\partial^l}{\partial \mathbf v^l}\frac{1}{v}+v^{2l+1}\frac{\partial^{l+1}}{\partial \mathbf v^{2l+1}}\frac{1}{v}\right]\nonumber\\
    &=\frac{2l+1}{v^2}\left[\mathbf v \frac{v^{2l+1}(-1)^l}{(2l-1)!!}\frac{\partial^l}{\partial \mathbf v^l}\frac{1}{v}-\frac{(-1)^{l+1}v^{2(l+1)+1}}{(2l+1)!!}\frac{\partial^{l+1}}{\partial \mathbf v^{l+1}}\frac{1}{v}\right]\nonumber\\
    &=\frac{2l+1}{v^2}\left[\mathbf v \mathbf Y^l(\mathbf v)-\mathbf Y^{l+1}(\mathbf v)\right],
\label{eq:recyl}
\end{align}
which is the same recursion relation present in \cref{eq:recpl}.
Since $\mathbf Y^{0}(\mathbf v)= \mathbf P^0(\mathbf v)$ and $\mathbf Y^{1}(\mathbf v)=\mathbf P^1(\mathbf v)$, the proof is complete.

The integrals in \cref{eq:roshb4} can be put in terms of upper
\begin{equation}
    I_+^k=\frac{1}{\sqrt{\pi}}\int_0^{x_b} dx e^{-x} x^{(k-1)/2},
\end{equation}
and lower
\begin{equation}
    I_-^k=\frac{1}{\sqrt{\pi}}\int_{x_b}^{\infty} dx e^{-x} x^{(k-1)/2},
\end{equation}
incomplete gamma functions \citep{Abramowitz1972}, yielding
\begin{equation}
    H_b = \frac{2n_b}{v_{thb}} \sum_{l,k} \sum_{m=0}^k \frac{L_{km}^{l}}{\sigma_k^l}\frac{\mathbf Y^{l}(\hat v) \cdot \mathbf M_b^{lk}}{2l+1} \left(\frac{I_+^{2l+2m+2}}{x_b^{(l+1)/2}}+x_b^{l/2}I_-^{2m+1}\right).
\label{eq:roshb5}
\end{equation}
A procedure similar to the one used to obtain \cref{eq:roshb5} can be followed for the second Rosenbluth potential $G_b$ by expanding the distribution function $f_b$ appearing in $G_b$ according to \cref{eq:faexp}, therefore obtaining
\begin{equation}
\begin{split}
    G_b &= \frac{2n_b}{v_{thb}} \sum_{l,k} \sum_{m=0}^k \frac{L_{km}^{l}}{\sigma_k^l}\frac{\mathbf Y^{l}(\hat v) \cdot \mathbf M_b^{lk}}{2l+1}\left[\frac{1}{2l+3} \left(\frac{I_+^{2l+2m+4}}{x_b^{(l+1)/2}}+x_b^{l/2+1}I_-^{2m+1}\right)\right.\\
    &\left.-\frac{1}{2l-1}\left(\frac{I_+^{2l+2m+2}}{x_b^{(l-1)/2}}+x_b^{l/2}I_-^{2m+3}\right)\right].
\end{split}
\label{eq:rosgb1}
\end{equation}

Having derived a closed-form expression for the Rosenbluth potentials, we now turn to the full Coulomb collision operator.
We first note that, although the Rosenbluth potentials $H_b$ and $G_b$ are linear functions of $f_b$, the Coulomb collision operator is, in fact, bilinear in $f_a$ and $f_b$.
In order to rewrite the Coulomb collision operator in \cref{eq:coulombop} in terms of a single spherical harmonic tensor $\mathbf Y^l(\mathbf v)$, we make use of the following identity between symmetric traceless tensors \citep{Ji2009}
\begin{equation}
    [\mathbf Y^{l-u}(\hat v) \cdot \mathbf M_a^{lk}]\cdot^u [\mathbf Y^{n-u}(\hat v) \cdot \mathbf M_a^{nk}]=\sum_{j=0}^{min(l,n)-u}d_j^{l-u,n-u}\mathbf Y^{l+n-2(j+u)}(\hat v) \cdot \left({\mathbf M_a^{lk} \cdot^{j+u} \mathbf M_b^{nq}}\right)_{TS},
\label{eq:identityyln}
\end{equation}
where $\cdot^n$ is the $n$-fold inner product [e.g., for the matrix $\mathbf A = A_{ij}$, $(\mathbf A \cdot^1 \mathbf A)_{ij} = \sum_k A_{ki}A_{kj}$].
The $d_j^{l,n}$ coefficient can be written in terms of the $t_j^{l,n}$ coefficient
\begin{equation}
    t_j^{l,n}=\frac{l!n!(-2)^j(2l+2n-2j)!(l+n)!}{(2l+2n)!j!(l-j)!(n-j)!(l+n-j)!},
\end{equation}
as
\begin{equation}
    d_j^{l,n}=\sum_{j_k|\sum_{k=1}^hj_k=j}(-1)^h\prod_{k=1}^h t_{j_k}^{l-\sum_{g=1}^{k-1}j_g,n-\sum_{g=1}^{k-1}j_g}.
\label{eq:djln}
\end{equation}
{In \cref{eq:djln}, the summation range involves the integer partitions of $j$, i.e., the decomposition of $j$ into different sums of $h$ positive integers, here labeled as $j_k$, with $k$ ranging from $1$ to $h$ (e.g. for $j=3$, we obtain for $h=2$ the terms $j_1=2$ and $j_2=1$, and for $h=3$ we obtain $j_1=j_2=j_3=1${)}.}
Expanding $f_a$ and $f_b$ using \cref{eq:faexp}, the expression for the Rosenbluth potentials in \cref{eq:roshb5,eq:rosgb1}, and the identity in \cref{eq:identityyln}, the collision operator in \cref{eq:coulombop} can be rewritten in terms of products of $\mathbf M_a^{lk}$ and $\mathbf M_b^{lk}$ as
\begin{equation}
\begin{split}
    C(f_a,f_b)&=f_{aM}\sum_{l,k,n,q=0}^\infty\sum_{m=0}^k\sum_{r=0}^{q}\frac{L_{km}^l}{\sigma_k^l}\frac{L_{qr}^n}{\sigma_q^n} c_{ab}^{lkmnqr},
\end{split}
\label{eq:JiCab}
\end{equation}
with
\begin{equation}
    c_{ab}^{lkmnqr}=\sum_{u=0}^{\text{min}(2,l,n)}\nu_{*abu}^{lm,nr}(v^2)\sum_{i=0}^{min(l,n)-u}d_i^{l-u,n-u}\mathbf Y^{l+n-2(i+u)}(\hat v) \cdot \left({\mathbf M_a^{lk} \cdot^{i+u} \mathbf M_b^{nq}}\right)_{TS}.
\label{eq:ccjiheld}
\end{equation}
The quantity $\nu_{*abu}^{lm,nr}$ consists of a linear combination of $I_+^l$ and $I_-^l$ integrals and its derivatives, which can be written as linear combinations of the error function and its derivatives.
Their expressions are reported in \citet{Ji2009}.

\section{Gyrokinetic Coulomb Collision Operator}
\label{sec:gudingcentertransf}

In \cref{sec:gcmodel}, the Coulomb collision operator is cast in terms of velocity moments of the multipole expansion of the particle distribution functions $f_a$ and $f_b$.
We now express it in terms of the gyrokinetic distribution functions $\left< \overline{F_a} \right>$ and $\left< \overline{F_b} \right>$.
As a first step, the gyroangle dependence of the basis functions $\mathbf Y^{lk}$ is found explicitly by using a coordinate transformation from the particle phase-space coordinates $(\mathbf x, \mathbf v)$  to the guiding-center coordinate system $\mathbf Z$.
This allows us to decouple the fast gyromotion time associated with the gyroangle $\theta$ from the typical turbulence time scales.
The multipole moments $\mathbf M_a^{lk}$ and $\mathbf M_b^{lk}$ can then be written in terms {of} moments of the guiding-center distribution function $\left< F_a \right>$ and $\left< F_b \right>$ for arbitrary values of $k_\perp \rho_s$.
As a second step, the gyrocenter coordinate system $\overline{\mathbf Z}$ is introduced by using the coordinate transformation $T$ in \cref{eq:gyrotransf}.
As shown in \cref{sec:ordering}, for a gyrokinetic equation up to second order accurate in $\epsilon_\delta$, only the lowest order collision operator $C_0$ needs to be retained.
This allows us to straightforwardly obtain the gyrokinetic collision operator from the guiding-center one by a simple coordinate relabeling.

We first derive the polar and azimuthal angle (gyroangle) dependence of the $\mathbf Y^l(\mathbf v)$ tensor in terms of scalar spherical harmonics.
This is useful to analytically perform the gyroaverage of the collision operator in the Boltzmann equation, \cref{eq:boltzmannfinal}.
For this purpose, as a first step, we show that the Laplacian of $\mathbf Y^l(\mathbf v)$ vanishes, i.e. that $\mathbf Y^l(\mathbf v)$ are harmonic tensors.
By applying the operator $\nabla_{\mathbf v}^2$ to $ \mathbf Y^{l}(\mathbf v)$, and recalling that $\nabla_{\mathbf v}^2 (1/v) = 0$ for $v\not=0$, we obtain
\begin{equation}
    \nabla_{\mathbf v}^2 \mathbf Y^{l}(\mathbf v) = \frac{2(-1)^l(2l+1)v^{2l+1}}{(2l-1)!!}\left[(l+1)\left(\frac{\partial}{\partial \mathbf v}\right)^l\frac{1}{v}+\mathbf v \cdot \left(\frac{\partial}{\partial \mathbf v}\right)^{l+1}\frac{1}{v}\right]=0,
\label{eq:laplyl}
\end{equation}
since
\begin{equation}
    \mathbf v \cdot \left(\frac{\partial}{\partial \mathbf v}\right)^{l+1}\frac{1}{v} = - (l+1)\left(\frac{\partial}{\partial \mathbf v}\right)^l\frac{1}{v},
\end{equation}
as can be proved by induction \citep{Weinert1980}.
The angular dependence of $\mathbf Y^{l}(\mathbf v)$ can now be found by expressing the Laplacian of \cref{eq:laplyl} in spherical coordinates.
Using the fact that $\mathbf Y^l(\mathbf v) = v^l \mathbf Y^l(\hat v)$, we obtain
\begin{align}
    0 &= \nabla_{\mathbf v}^2 \mathbf Y^{l}(\mathbf v)=\nabla^2_{\mathbf v}[v^l \mathbf Y^l(\hat v)]\nonumber\\
      &= \mathbf Y^{l}(\hat v)\left(\frac{\partial^2}{\partial v^2} + \frac{2}{v}\frac{\partial}{\partial v}\right)v^l - v^{l-2} L^2 \mathbf Y^l(\hat v),
\label{eq:nabla2yl}
\end{align}
where $L^2$ is the angular part of the operator $\nabla^2_{\mathbf v}$ multiplied by $v^2$
\begin{equation}
    L^2 = \frac{1}{\sin \varphi}\frac{\partial}{\partial \varphi}\left(\sin \varphi \frac{\partial}{\partial \varphi}\right)+\frac{1}{\sin \varphi^2}\frac{\partial^2}{\partial \theta^2},
\end{equation}
with $\varphi$ and $\theta$ chosen, respectively, as the pitch angle and the gyroangle variables, both defined in \cref{eq:gcv}.
Evaluating the $v$ derivatives in \cref{eq:nabla2yl}, the following differential equation for $\mathbf Y^l(\mathbf v)$ is obtained
\begin{equation}
    L^2 \mathbf Y^l(\hat v) = l(l+1) \mathbf Y^l(\hat v).
\label{eq:l2yl}
\end{equation}
We identify \cref{eq:l2yl} as the eigenvalue equation for the scalar spherical harmonics $Y_{lm}(\varphi, \theta)$ \citep{Arfken2013}, which can be written in terms of associated Legendre polynomials $P_{l}^m(\cos \varphi)$ as \citep{Abramowitz1972}
\begin{equation}
    Y_{lm}(\varphi,\theta) = (-1)^m\sqrt{\frac{(2l+1)}{4 \pi}\frac{(l-m)!}{(l+m)!}}P_{l}^m(\cos \varphi)e^{i m \theta}.
\label{eq:ylmasslag}
\end{equation}
with
\begin{equation}
    P_l^m(x)=(1-x^2)^{m/2}\frac{d^m}{dx^m}[P_l(x)],
\end{equation}
and $P_l(x)=(d^l/dx^l)[(x^2-1)^l]/(2^l l!)$ a Legendre polynomial.
%
%
Therefore, using \cref{eq:l2yl}, and denoting $\mathbf e^{l m}$  the basis elements of $\mathbf Y^l(\mathbf v)$ (an elementary derivation of the basis tensors $\mathbf e^{l m}$ is shown in \cref{app:basistensors}), we write $\mathbf Y^{l}(\mathbf v)$ as
\begin{equation}
    \mathbf Y^{l}(\mathbf v) = v^l \sqrt{\frac{ 2 \pi^{3/2}  l!}{2^l (l+1/2)!}}\sum_{m=-l}^lY_{lm}(\varphi,\theta)\mathbf e^{l m}.
\label{eq:ylvfull}
\end{equation}

Having derived the gyroangle dependence of the $\mathbf Y^{l}(\mathbf v)$ tensors, we now compute the fluid moments $\mathbf M_a^{lk}$ in terms of $v_\parallel$ and $\mu$ moments of the guiding-center distribution function $\left<F_a\right>$.
In order to perform the velocity integration in the definition of the moments $\mathbf M^{lk}$ in \cref{eq:defmaln} at arbitrary $k_\perp \rho$ in guiding-center phase-space coordinates, we use the identity $f(\mathbf x,\mathbf v) = \int f(\mathbf x',\mathbf v)\delta(\mathbf x-\mathbf x')d\mathbf x'$.
By imposing $\mathbf x' = \mathbf R+\mathbf \rho$, writing the volume element in phase-space as $d\mathbf x'd\mathbf v = (B_\parallel^*/m)d\mathbf R dv_\parallel d\mu d\theta$, and using \cref{eq:gdpf}, we obtain
\begin{equation}
    n_a \mathbf M_a^{lk}(\mathbf x)= \int F_a(\mathbf R, v_\parallel, \mu, \theta) \mathbf Y^{lk}\left(\frac{\mathbf v}{v_{tha}}\right) \delta(\mathbf x - \mathbf R - \mathbf \rho_a) \frac{B_{\parallel}^*}{m}d\mathbf R dv_\parallel d\mu d\theta.
\label{eq:gcmalndelta}
\end{equation}
from \cref{eq:defmaln}.
Expressing $\mathbf v = \mathbf v(\mathbf Z)$, as shown by \cref{eq:gcv}, and performing the integral over $\mathbf R$ in \cref{eq:gcmalndelta}, it follows that
\begin{equation}
    n_a \mathbf M_a^{lk}(\mathbf x)= \int F_a(\mathbf x-{\mathbf \rho}_a, v_\parallel, \mu,\theta) \mathbf Y^{lk}\left[\frac{\mathbf v(\mathbf x-{\mathbf \rho}_a, v_\parallel, \mu, \theta)}{v_{tha}}\right] \frac{B_{\parallel}^*}{m}dv_\parallel d\mu d\theta.
\label{eq:gcmalnnodelta}
\end{equation}
The orderings in \cref{eq:orderingftildee,eq:orderingftildei} for the guiding-center distribution function $F_a$ allows us to approximate $F_a \simeq \left< F_a \right>_{\mathbf R}$ \citep{Jorge2017}, effectively neglecting $\epsilon^2$ effects in $\mathbf M_a^{lk}$, hence in the collision operator $C(f_a,f_b)$.
To make further analytical progress, {and in line with previous gyrokinetic literature \citep{Li2011, Pan2019},} we represent $F_a(\mathbf R, v_\parallel, \mu, \theta)$ by its Fourier transform $F_a(\mathbf k, v_\parallel, \mu, \theta)=\int F_a(\mathbf R, v_\parallel, \mu, \theta) e^{-i \mathbf k \cdot \mathbf R} d \mathbf R$, and write
\begin{equation}
    n_a \mathbf M_a^{lk}(\mathbf x)= \int \left<F_a(\mathbf k, v_\parallel, \mu,\theta) \right>_{\mathbf R}\mathbf Y^{lk}\left[\frac{\mathbf v(\mathbf x-{\mathbf \rho}_a, v_\parallel, \mu, \theta)}{v_{tha}}\right] e^{i \mathbf k \cdot \mathbf x}e^{-i \mathbf k \cdot \rho_a} \frac{B_{\parallel}^*}{m}d \mathbf k dv_\parallel d\mu d\theta.
\label{eq:gcmalnnodelta2}
\end{equation}
By aligning the $\mathbf k$ coordinate system in the integral of \cref{eq:gcmalnnodelta2} with the axes $(\mathbf b, \mathbf e_1, \mathbf e_2)$, i.e. $\mathbf k = k_\parallel \mathbf b + k_\perp(\cos \theta \mathbf e_1 + \sin \theta \mathbf e_2)$, we write $\exp(-i \mathbf k \cdot \mathbf \rho)=\exp(-i k_\perp \rho \cos \theta)$.
We then use the Jacobi-Anger expansion \citep{Andrews1992}
\begin{equation}
    e^{-i k_\perp \rho \cos \theta}=\sum_{p=-\infty}^{\infty}(-i)^p J_p(k_\perp \rho)e^{- i p \theta},
\label{eq:jacoan}
\end{equation}
with $J_p$ the Bessel function of order $p$, and rewrite \cref{eq:gcmalnnodelta2} as
\begin{equation}
\begin{split}
    n_a \mathbf M_a^{lk}(\mathbf x)= \sum_{p=-\infty}^{\infty}(-1)^p &\int J_p(k_\perp \rho)\left<F_a(\mathbf k, v_\parallel, \mu,\theta)\right>_{\mathbf R}  e^{i \mathbf k \cdot \mathbf x}\\
    &\times\mathbf Y^{lk}\left[\frac{\mathbf v(\mathbf x-{\mathbf \rho}_a, v_\parallel, \mu, \theta)}{v_{tha}}\right] e^{-ip\theta} \frac{B_{\parallel}^*}{m}d \mathbf k dv_\parallel d\mu d\theta.
\end{split}
\label{eq:gcmalnnodelta3}
\end{equation}
The velocity $\mathbf v$ in the argument of $\mathbf Y^{lk}$ in \cref{eq:gcmalnnodelta3} is then expanded as
\begin{equation}
    \mathbf v(\mathbf x-{\mathbf \rho}_a, v_\parallel, \mu, \theta) = \mathbf v(\mathbf x, v_\parallel, \mu, \theta) + O(\epsilon_B).
\label{eq:approxvR}
\end{equation}
The second term in \cref{eq:approxvR} introduces {$\epsilon_B \ll \epsilon$} terms in the collision operator and is therefore neglected.
{An example of a numerical implementation using a similar Fourier representation can be found in \citet{Pan2019}.}

Using \cref{eq:ylvfull} to express $\mathbf Y^{lk}(\mathbf v)$ in terms of spherical harmonics, we perform the gyroangle integration in \cref{eq:gcmalnnodelta3}.
By rewriting the spherical harmonics $Y_{lm}(\varphi,0)$ in terms of associated Legendre polynomials $P_{l}^m(\cos \varphi)$ using \cref{eq:ylmasslag}, the gyroaverage of the product $Y^{lk}(\mathbf v/v_{tha}) e^{-ip\theta}$ can be performed, yielding
\begin{equation}
\begin{split}
    n_a \mathbf M_a^{lk}(\mathbf x)= \sum_{p=-\infty}^{\infty}(-1)^p &\int J_p(k_\perp \rho_a)\left<F_a(\mathbf k, v_\parallel, \mu,\theta)\right>_{\mathbf R}  e^{i \mathbf k \cdot \mathbf x}\\
    &\times\left<\mathbf Y^{lk}\left(\frac{\mathbf v}{v_{tha}}\right) e^{-ip\theta}\right> \frac{B_{\parallel}^*}{m}d \mathbf k dv_\parallel d\mu 2\pi,
\end{split}
\label{eq:gcmalnnodelta4}
\end{equation}
with
\begin{equation}
\begin{split}
    \left<\mathbf Y^{lk}\left(\frac{\mathbf v}{v_{tha}}\right) e^{-ip\theta}\right>&={L_k^{l+1/2}\left(\frac{v}{v_{tha}}\right)}\left(\frac{v}{v_{tha}}\right)^l\sqrt{\frac{\pi^{1/2}l!}{2^l(l-1/2)!}}\\
    &\times\sum_{m=-l}^l (-1)^m \sqrt{\frac{(l-m)!}{(l+m)!}} P_l^m(\cos {\varphi}){\mathbf e^{lm}}\delta_{m{,}p}.
\label{eq:generalizedjiheld}
\end{split}
\end{equation}
We note that, the $p=0$ case of \cref{eq:generalizedjiheld} corresponds to the gyroaveraged formulas in \citet{Ji2009a,Ji2013a,Ji2014a} used to derive closures for fluid models {at zeroth order in $\epsilon$}.
Finally, by defining the Bessel-Fourier operator
\begin{equation}
     j_m[F_a]\equiv \int J_m(k_\perp \rho_a)\left<F_a(\mathbf k, v_\parallel, \mu,\theta)\right>_{\mathbf R}  e^{i \mathbf k \cdot \mathbf x}d \mathbf k,
\label{eq:fourierbessel}
\end{equation}
the expression for the fluid moments $\mathbf M_a^{lk}$ in terms of coupled $v_\parallel$ and $\mu$ moments of the guiding-center distribution function $\left<F_a\right>_{\mathbf R}$ is obtained
\begin{equation}
\begin{split}
    n_a \mathbf M_a^{lk}(\mathbf x)&= \sqrt{\frac{2\pi^{5/2}l!}{2^l(l+1/2)!}}\sum_{m=-l}^{l} \mathbf e^{l m}(-1)^m \mathcal{M}^{lk}_{am}(\mathbf x),
\end{split}
\label{eq:gcmalnfinal}
\end{equation}
with
\begin{equation}
    \mathcal{M}^{lk}_{am}(\mathbf x)=\int j_m[F_a] v^l L_k^{l+1/2}(x_a^2) Y_{lm}(\varphi,0)
     \frac{B_{\parallel}^*}{m} dv_\parallel d\mu.
\end{equation}

Equation (\ref{eq:gcmalnfinal}) can now be used to express the collision operator $C(f_a,f_b)$ in terms of $v_\parallel$ and $\mu$ integrals of $\left< F_a \right>$.
Using \cref{eq:ccjiheld,eq:ylvfull,eq:JiCab}, and defining
\begin{equation}
    E_{jv}^{ls nt} = \mathbf e^{l+n-2j~v}\cdot ({\mathbf e^{ls}\cdot^{j}\mathbf e^{nt}})_{TS},
\end{equation}
we can write the collision operator in \cref{eq:JiCab,eq:ccjiheld} as a function of the $\mathcal{M}^{lk}_{am}$ moments, i.e.
\begin{equation}
\begin{split}
    c^{lkmnqr}_{ab}&=\sum_{u=0}^{\text{min}(2,l,n)}\sum_{j=0}^{\text{min}(l,n)-u}d_j^{l-u,n-u}a^{ln}_{j+u}\sum_{s=-l}^l\sum_{t=-n}^n\sum_{v=-(l+n-2j-2u)}^{l+n-2j-2u} E^{ls nt}_{j+u~v}\\
    &\times  Y_{l+n-2j-2u~v}(\varphi,\theta)\frac{\nu_{*abu}^{lm,nr}(v^2)}{n_a n_b} \mathcal{M}^{lk}_{as}(\mathbf x)\mathcal{M}^{nq}_{bt}(\mathbf x),
\end{split}
    \label{eq:ccgc}
\end{equation}
with
\begin{equation}
    a^{ln}_{j}=\frac{1}{2^{l+n-j}}\sqrt{\frac{8 \pi^{13/2}l!n!(l+n-2j)!}{(l+1/2)!(n+1/2)!(l+n-2j+1/2)!}}.
\end{equation}

We now focus on the gyroaverage of the collision operator in \cref{eq:ccgc} with the gyroaverage operation performed at constant $\mathbf R$.
We first note that the gyroangle $\theta$ dependence in $c^{lkmnqr}_{ab}$ is present through the spherical harmonic $ Y_{l+n-2j-2u~v}(\varphi,\theta)$ and through the fluid moments $\mathcal{M}^{lk}_{as}$ and $\mathcal{M}^{nq}_{bt}$ as the latter are functions of $\mathbf x = \mathbf R +{\mathbf \rho_a}$.
To make the gyroangle dependence explicit, we write both $\mathcal{M}^{lk}_{as}$ and $\mathcal{M}^{nq}_{bt}$ in Fourier space as
\begin{equation}
    \mathcal{M}^{lk}_{as}(\mathbf x)\mathcal{M}^{nq}_{bt}(\mathbf x) = \int d\mathbf k d \mathbf k' e^{i(\mathbf k + \mathbf k')\cdot \mathbf R}\mathcal{M}^{lk}_{as}(\mathbf k)\mathcal{M}^{nq}_{bt}(\mathbf k')e^{i(\mathbf k \cdot \mathbf \rho_a + \mathbf k'\cdot {\mathbf \rho_a})}.
\label{eq:Fouriermoment}
\end{equation}
Using the Jacobi-Anger expansion of \cref{eq:jacoan}, we find that
\begin{align}
    &\left<Y_{lm}(\varphi,\theta)\mathcal{M}^{lk}_{as}(\mathbf x)\mathcal{M}^{nq}_{bt}(\mathbf x)\right>_{\mathbf R}=\int d\mathbf k d \mathbf k' e^{i(\mathbf k + \mathbf k')\cdot \mathbf R}\mathcal{M}^{lk}_{as}(\mathbf k)\mathcal{M}^{nq}_{bt}(\mathbf k') \nonumber\\
    &\times\sqrt{\frac{2l+1}{4\pi}\frac{(l-m)!}{(l+m)!}}P_l^m(\cos \varphi){\sum_{p={-\infty}}^{\infty}(-1)^{p+m}e^{i (p+m)\alpha}J_p(k_\perp \rho_a)J_{p+m}(k_\perp' {\rho_a})},
\end{align}
{with $\alpha$ the azimuthal angle of the $\mathbf k'$ vector, i.e., the angle between $\mathbf k_\perp'$ and {$\mathbf k_\perp$}.}
The gyroaveraged collision operator at arbitrary $k_\perp \rho$ is therefore given by
\begin{equation}
    \left<C(F_a,F_b)\right>_{\mathbf R}=f_{aM}\sum_{l,k,n,q=0}^\infty\sum_{m=0}^k\sum_{r=0}^{q}{L_{km}^lL_{qr}^n} \left<c^{lkmnqr}_{ab}\right>_{\mathbf R},
    \label{eq:JiCabgyro}
\end{equation}
with
\begin{align}
    \left<c^{lkmnqr}_{ab}\right>_{\mathbf R}&=\sum_{u=0}^{\text{min}(2,l,n)}\sum_{j=0}^{\text{min}(l,n)-u}d_j^{l-u,n-u}a^{ln}_{j+u}\sum_{s=-l}^l\sum_{t=-n}^n \sum_{v=-(l+n-2j-2u)}^{l+n-2j-2u} E^{ls nt}_{j+u~v}\nonumber\\
    &\times b_{j+u}^{l+n v} P_{l+n-2j-2u}^v(\cos \varphi){\nu_{*abu}^{lm,nr}(v^2)}  \int \mathcal{M}^{lk}_{as}(\mathbf k)\mathcal{M}^{nq}_{bt}(\mathbf k')e^{i(\mathbf k +\mathbf k')\cdot \mathbf R} \nonumber\\
    &\times
    {\sum_{p={-\infty}}^{\infty}(-1)^{p+v}e^{i (p+v)\alpha}J_p(k_\perp \rho_a)J_{p+v}(k_\perp' {\rho_a})}
    d \mathbf k d\mathbf k',
    \label{eq:ccgcgyro}
\end{align}
and
\begin{equation}
    b_{j}^{lv}=i^v\sqrt{\frac{2l-4j}{4\pi}\frac{(l-2j-v)!}{(l-2j+v)!}}
\end{equation}
We note that, if only first order $k_\perp \rho$ terms are kept in the Fourier-Bessel operator of \cref{eq:fourierbessel}, the collision operator in \cref{eq:JiCabgyro} reduces to the drift-kinetic collision operator found in \citet{Jorge2017}.

In \cref{eq:JiCabgyro}, the gyroaveraged collision operator is cast in terms of $v_\parallel$ and $\mu$ moments of the guiding-center distribution function $\left< F_a \right>$ for arbitrary values of $k_\perp \rho$.
We now apply the transformation $T$, introduced in \cref{eq:gyrotransf}, to \cref{eq:JiCabgyro} in order to write the gyroaveraged collision operator in terms of $\overline v_\parallel$ and $\overline \mu$ moments of the gyrocenter distribution function $\left<\overline{F_a} \right>_{\overline{\mathbf R}}$.
As shown in \cref{sec:ordering}, only the zeroth order terms in the $\epsilon_\delta$ expansion of $\left<C(\overline{F_a}, \overline{F_b})\right>_{\overline{\mathbf R}}$ are needed in order to adequately describe collisional processes in the gyrokinetic framework.
Therefore, using \cref{eq:gyrotransf}, we apply the zeroth order transformations $\mathbf Z \simeq \overline{\mathbf Z}$ and $F_a(\mathbf Z) = T \overline{F_a}(\mathbf Z) \simeq \overline{F_a}(\mathbf Z)$ to the collision operator $\left< C(F_a,F_b) \right>$ in \cref{eq:JiCabgyro}, yielding
\begin{equation}
    \left<C(\overline{F_a},\overline{F_b})\right>_{\overline {\mathbf R}}\simeq f_{aM}\sum_{l,k,n,q=0}^\infty\sum_{m=0}^k\sum_{r=0}^{q}{L_{km}^lL_{qr}^n} \left<\overline c^{lkmnqr}_{ab}\right>_{\overline {\mathbf R}},
    \label{eq:JiCabgyrok}
\end{equation}
with
\begin{align}
    \left<\overline c^{lkmnqr}_{ab}\right>_{\overline {\mathbf R}}&=\sum_{u=0}^{\text{min}(2,l,n)}\sum_{j=0}^{\text{min}(l,n)-u}d_j^{l-u,n-u}a^{ln}_{j+u}\sum_{s=-l}^l\sum_{t=-n}^n \sum_{v=-(l+n-2j-2u)}^{l+n-2j-2u} E^{ls nt}_{j+u~v}\nonumber\\
    &\times b_{j+u}^{l+n v}P_{l+n-2j-2u}^v(\overline v_\parallel/\overline v){\nu_{*abu}^{lm,nr}(\overline v^2)} \int \mathcal{\overline M}^{lk}_{as}(\mathbf k)\mathcal{\overline M}^{nq}_{bt}(\mathbf k')e^{i(\mathbf k +\mathbf k')\cdot \overline{\mathbf{R}}} \nonumber\\
    &\times
    {\sum_{p={-\infty}}^{\infty}(-1)^{p+v}e^{i (p+v)\alpha}J_p(k_\perp \overline \rho_a)J_{p+v}(k_\perp' {\overline \rho_a})}
    d \mathbf k d\mathbf k',
    \label{eq:ccgcgyrok}
\end{align}
where $\overline{\rho_a}=\rho_a(\overline{\mathbf R},\overline \mu, \overline \theta)$, $\overline v^2 = \overline v_\parallel^2+2 B \overline \mu/m$, and
\begin{equation}
    \overline{\mathcal{M}}^{lk}_{am}=\int j_m[\overline{F_a}] \overline v^l L_k^{l+1/2}(\overline v^2) Y_{lm}\left(\overline \varphi,0\right)\frac{B_{\parallel}^*}{m} d \overline v_\parallel d\overline \mu.
\label{eq:gyrokma}
\end{equation}
The collision operator in \cref{eq:JiCabgyrok} represents the gyrokinetic full Coulomb collision operator that can be used in gyrokinetic models that are up to $O(\epsilon_\delta^2)$ accurate.
In \cref{eq:JiCabgyrok}, the integral-differential character of the $C(f_a,f_b)$ operator is replaced by two-dimensional integrals of the gyrocenter distribution function over the velocity coordinates $\overline v_\parallel$ and $\overline \mu$.
{We note that, in practice, a truncation of the series present in \cref{eq:JiCabgyrok,eq:ccgcgyrok} requires a numerical study in order to assess their convergence.
}

\section{Hermite-Laguerre Expansion of the Coulomb Operator}
\label{eq:hermlag}
In this section, we expand the distribution function into an orthogonal Hermite-Laguerre polynomial basis and compute the Hermite-Laguerre moments of the Coulomb collision operator in \cref{eq:JiCabgyrok}.
An expansion of the drift-kinetic \citep{Jorge2017} and gyrokinetic \citep{Mandell2018,Frei2019} equation in Hermite-Laguerre polynomials has been recently introduced, showing that this is an advantageous approach to the study of plasma waves and instabilities \citep{Jorge2018,Jorge2019}.
A key reason for using a basis of Hermite-Laguerre polynomials in gyrokinetics is that these polynomials are orthogonal with respect to a Maxwellian, and can be directly related to the Bessel functions used in evaluating gyroaverage operators such as the ones present in \cref{eq:ccgcgyrok,eq:gyrokma}.
We therefore expand $\left< \overline{F_a} \right>_{\overline {\mathbf R}}$ as
\begin{equation}
    \left< \overline{F_a} \right>_{\overline {\mathbf R}} = f_{Ma}\sum_{p,j}\frac{\overline N_a^{pj}}{\sqrt{2^p p!}}H_p\left(\overline s_{\parallel a}\right)L_j\left(\overline s_{\perp a}^2\right),
\label{eq:gyrof}
\end{equation}
where $H_p$ are {\textit{physicists'}} Hermite polynomials of order $p$ defined by the Rodrigues' formula
\begin{equation}
    H_p(x)=(-1)^p e^{x^2}\frac{d^p}{dx^p}e^{-x^2},
\end{equation}
and normalized via
\begin{equation}
    \int_{-\infty}^{\infty} dx H_p(x) H_{p'}(x) e^{-x^2} = 2^p p! \sqrt{\pi} \delta_{p{p'}},
\end{equation}
and $L_j$ the Laguerre polynomials of order $j$ defined by the Rodrigues' formula
\begin{equation}
    L_j(x)=\frac{e^x}{j!}\frac{d^j}{dx^j}(e^{-x}x^j),
\end{equation}
and orthonormal with respect to the weight $e^{-x}$
\begin{equation}
    \int_{0}^{\infty} dx L_j(x) L_{j'}(x) e^{-x} = \delta_{jj'}.
\end{equation}
In \cref{eq:gyrof}, we introduce the normalized parallel velocity
\begin{equation}
    \overline s_{\parallel a}=\frac{\overline v_{\parallel}}{v_{tha}},
\end{equation}
and the perpendicular velocity coordinate
\begin{equation}
    \overline s_{\perp a}^2=\frac{\overline \mu B}{T_a}.
\end{equation}
Due to the orthogonality of the Hermite-Laguerre polynomial basis, the coefficients $N_a^{pj}$ of the expansion in \cref{eq:gyrof} can be computed as
\begin{equation}
    \overline N_a^{pj}=\int \frac{H_p\left(\overline s_{\parallel a}\right)L_j\left(\overline s_{\perp a}^2\right) \left< \overline{F_a} \right>_{\overline {\mathbf R}} }{\sqrt{2^p p!}} \frac{B}{m_a} d\overline v_\parallel d \overline \mu d \overline \theta.
    \label{eq:gyromoments}
\end{equation}
We note that the integrand of $\overline N_a^{pj}$ in \cref{eq:gyromoments} contains the multiplicative factor $B/m_a$, as opposed to the Jacobian containing the factor $B_\parallel^*/m_a$.
In the following, we also use the Hermite-Laguerre moments of $\left< \overline{F_a} \right>_{\overline {\mathbf R}}$ with $B_\parallel^*$ in the integrand instead of B.
These are denoted as $\overline N_a^{*pj}$, i.e.
\begin{equation}
\begin{split}
    \overline N_a^{*pj}&=\int \frac{H_p\left(\overline s_{\parallel a}\right)L_j\left(\overline s_{\perp a}^2\right) \left< \overline{F_a} \right>_{\overline {\mathbf R}} }{\sqrt{2^p p!}} \frac{B_\parallel^*}{m_a} d\overline v_\parallel d \overline \mu d \overline \theta\\
    &=\overline N_a^{pj}\left(1+\frac{\mathbf b \cdot \nabla \times \mathbf v_{\mathbf E}}{\Omega_a}\right)+ \frac{v_{tha}}{\sqrt{2}\Omega_a}\mathbf b \cdot \nabla \times \mathbf b\left(\sqrt{p+1}\overline N_a^{p+1~j}+\sqrt{p}\overline N_a^{p-1~j}\right),
\end{split}
    \label{eq:gyromomentsstar}
\end{equation}
while in the weak flow regime the term proportional to $\mathbf b \cdot \nabla \times \mathbf v_{\mathbf E}/\Omega_a$ in \cref{eq:gyromomentsstar} is set to zero.

In order to express the collision operator in terms of the moments $\overline N_a^{pj}$ given in \cref{eq:gyromoments} and evaluate its Hermite-Laguerre moments, we first consider the gyrokinetic moments $\overline{\mathcal{M}}_{am}^{lk}$ and write the integral that defines them in \cref{eq:gyrokma} as a function of the gyrocenter moments $\overline N_a^{pj}$ of \cref{eq:gyromoments}.
As a first step, we project both the Fourier-Bessel operator $j_m[\overline{F_a}]$ and the spherical harmonics $Y_{lm}$ on the Hermite-Laguerre basis.
We remark that the $\overline \mu$ and $k_\perp$ dependence in the Fourier-Bessel operator $j_m$, \cref{eq:fourierbessel}, can be decomposed by introducing $\rho_{tha} = v_{tha}/\Omega_a$ and noting that $|\mathbf \rho_a| = \sqrt{\overline \mu B/T_a} \rho_{tha}=\overline s_{\perp a} \rho_{tha}$.
This allows the use of the following identity between Bessel and Legendre functions \citep{Zwillinger2014}
\begin{equation}
    J_m(2 b_a \overline s_{\perp a}) = \sigma_m b_a^{|m|} \overline s_{\perp a}^{|m|}e^{-b_a^2}\sum_{r=0}^\infty \frac{L_r^{|m|}(\overline s_{\perp a}^2)}{(|m|+r)!}b_a^{2r}.
\label{eq:bessLeg}
\end{equation}
with $b_a=k_\perp \rho_{tha}/2$, $\sigma_0=1$ and $\sigma_m = \text{sgn}(m)^m$ for $m\not=0$.
The Fourier-Bessel operator in \cref{eq:fourierbessel}, with the identity in \cref{eq:bessLeg} and the Hermite-Laguerre expansion of \cref{eq:gyrof}, can then be written as
\begin{equation}
    j_m[\overline{F_a}]=f_{Ma}\sum_{p=0}^\infty \sum_{j=0}^\infty \sum_{r=0}^\infty \frac{H_p(\overline s_{\parallel a}) L_j(\overline s_{\perp a}^2)}{\sqrt{2^p p!}} \frac{L_r^m( \overline s_{\perp a}^2) \overline s_{\perp a}^m}{(m+r)!}\int \overline N_a^{pj}(\mathbf k)b_a^{m+2r}e^{-b_a^2} e^{i \mathbf k \cdot \mathbf x} d\mathbf k.
\label{eq:jmhplj}
\end{equation}
As a second step, we consider
\begin{equation}
    Y_{lm}(\varphi,0)=(-1)^m\sqrt{\frac{2l+1}{4\pi}\frac{(l-m)!}{(l+m)!}}P_l^m(\cos \varphi).
\label{eq:ylm0}
\end{equation}
which are used in the definition in \cref{eq:ylmasslag}.
In order to expand the associated Legendre polynomials $P_l^m(\cos \varphi)$ appearing in \cref{eq:ylm0} in a Hermite-Laguerre basis, we generalize the basis transformation between a Legendre-Associated Laguerre and a Hermite-Laguerre basis presented in \citet{Jorge2017} to a transformation between an Associated Legendre-Associated Laguerre and a Hermite-Laguerre basis, that is
\begin{equation}
    \frac{\overline v^l}{v_{tha}^l} P_l^m\left(\frac{\overline v_\parallel}{\overline v}\right) L_k^{l+1/2}\left(\frac{\overline v^2}{v_{tha}^2}\right) = \sum_{p=0}^{l+2k}\sum_{j=0}^{k+\floor{l/2}}T_{lkm}^{pj}H_p\left(\frac{\overline v_{\parallel a}}{v_{tha}}\right)L_j\left(\frac{\overline \mu B}{T_a}\right)\left(\frac{\overline \mu B}{T_a}\right)^{m/2}.
\label{eq:plljhplj}
\end{equation}
For the derivation and expression of the $T_{lkm}^{pj}$ coefficients, see \cref{app:tlkpj1}.
The inverse transformation coefficients $\left(T^{-1}\right)_{pj}^{lkm}$ are defined as
\begin{equation}
    H_p\left(\frac{\overline v_{\parallel a}}{v_{tha}}\right)L_j\left(\frac{\overline \mu B}{T_a}\right)\left(\frac{\overline \mu B}{T_a}\right)^{m/2}=\sum_{l=0}^{p+2j}\sum_{k=0}^{j+\floor{p/2}}\left(T^{-1}\right)_{pj}^{lkm}\frac{\overline v^l}{v_{tha}^l} P_l^m\left(\frac{\overline v_\parallel}{\overline v}\right) L_k^{l+1/2}\left(\frac{\overline v^2}{v_{tha}^2}\right).
\label{eq:plljhpljminus1}
\end{equation}
The gyrocenter moments $\overline{\mathcal{M}}^{lk}_{am}$ in \cref{eq:gyrokma} can then be rewritten using the identities in \cref{eq:jmhplj,eq:plljhplj} and
\begin{equation}
    L_r^m(x)L_j(x)x^m = \sum_{s=0}^{m+r+j}d^m_{rjs} L_s(x),
\label{eq:doubleLaguerre}
\end{equation}
with the $d^r_{mjs}$ coefficients given by
\begin{equation}
    d^m_{rjs}=\sum_{r_1=0}^r \sum_{j_1=0}^j \sum_{s_1=0}^s L_{r r_1}^{-1/2} L_{j j_1}^{m-1/2} L_{s s_1}^{-1/2} (r_1+j_1+s_1+m)!,
\end{equation}
yielding the following expression
\begin{align}
    \overline{\mathcal{M}}_{am}^{lk}(\mathbf k)&=\sum_{g=0}^\infty\sum_{h=0}^{l+2k}\sum_{u=0}^{k+\floor{l/2}}\sum_{s=0}^{m+r+u}M_{lkmg}^{hus} \overline N_{a}^{*hs}(\mathbf k)b_a^{2g+m}e^{-b_a^2}.
\label{eq:gyromomentsfinal}
\end{align}
where we defined
\begin{equation}
    M_{lkmg}^{hus}=(-1)^m\frac{T_{lkm}^{hu} d_{gus}^m \sqrt{2^p p!}}{(m+g)!}\sqrt{\frac{2l+1}{4\pi}\frac{(l-m)!}{(l+m)!}}.
\end{equation}
Using the form for $\overline{\mathcal{M}}_{am}^{lk}$ in \cref{eq:gyromomentsfinal}, the collision operator in \cref{eq:JiCabgyrok} can therefore be expressed in terms of Hermite-Laguerre moments $N^{pj}$ of the distribution function.
We note that the moments $ \overline{\mathcal{M}}_{am}^{lk}$ in \cref{eq:gyromomentsfinal} reduce to the ones in \citet{Jorge2017} in the lowest order drift-kinetic limit $k_\perp \rho_{tha}=0$.

We now take Hermite-Laguerre moments of the collision operator $\left<C(\overline{F_a},\overline{F_b})\right>$, i.e. we evaluate
\begin{equation}
\begin{split}
    C_{ab}^{pj}(\overline{\mathbf R})&=\int \left<C(\overline{F_a},\overline{F_b})\right>_{\overline{\mathbf R}} \frac{H_p(\overline s_{\parallel a})L_j(\overline s_{\perp a^2})}{\sqrt{2^p p!}}\frac{B_\parallel^*}{m_a}d \overline v_\parallel d \overline \mu d\overline \theta.
\end{split}
\label{eq:collopmoments}
\end{equation}

Writing the gyroaveraged collision operator $\left<C(\overline{F_a},\overline{F_b})\right>$ in \cref{eq:JiCabgyrok} using \cref{eq:gyromomentsfinal,eq:ccgcgyrok}, and expanding the Bessel functions {$J_p(k_\perp \rho_a)$ and $J_{p+m}(k_\perp' {\rho_a})$} using \cref{eq:bessLeg}, the following form for the $\left<\overline c_{ab}^{lkmnqr} \right>_{\overline {\mathbf R}}$ term appearing in $\left<C(\overline{F_a},\overline{F_b})\right>_{\overline{\mathbf R}}$ is obtained
%
\begin{align}
    \left<\overline c_{ab}^{lkmnqr} \right>_{\overline {\mathbf R}}&=\sum_{u=0}^{min(2,l,n)}\sum_{i=0}^{min(l,n)-u} \sum_{d=-l-n+2i+2u}^{l+n-2i-2u}\sum_{z=0}^\infty{\sum_{p,p'=0}^\infty} \int D_{abuidz{pp'}}^{lkmnqr}(\mathbf k, \mathbf k')\nonumber\\
    &\times  P_{l+n-2i-2u}^d\left(\frac{\overline v_\parallel}{\overline v}\right){\overline s_{\perp a}^{d+2z}} {L_p^z(\overline s_{\perp a}^2)L_{p'}^{z+d}(\overline s_{\perp a}^2)} \nu_{*abu}^{lm,nr}(\overline v^2)e^{i (\mathbf k + \mathbf k')\mathbf R} d\mathbf k d\mathbf k'.
\label{eq:collophl}
\end{align}
In \cref{eq:collophl}, we defined the $D_{abuidz{pp'}}^{lkmnqr}$ term
\begin{equation}
    D_{abuidz{pp'}}^{lkmnqr}(\mathbf k, \mathbf k')=\sum_{s=-l}^l \sum_{t=-n}^n E_{i+ud}^{lsnt} {B_{a}^{zdp'}}\frac{d_i^{l-u,n-u}a_{i+u}^{ln}{e^{-b_a^2-b_{{a}}^{'2}}}}{{(p+z)!(z+d+p')!}}\mathcal{N}_{abuidz}^{lkmnqr}(\mathbf k, \mathbf k'),
\end{equation}
with {$B_{a}^{pvz'}=b_a^{p+2z}b_{{a}}^{p+v+2z'}$ and $b_{{a}}'=k_\perp'\rho_{th{a}}/2$}, while the convolution operator $\mathcal{N}_{abuidz}^{lkmnqr}(\mathbf k, \mathbf k')$ is given by
\begin{align}
    \mathcal{N}_{abuidz}^{lkmnqr}(\mathbf k, \mathbf k')&={(-1)^{z+d}e^{i(z+d)\alpha}}{b_{i+u}^{l+n d}} \overline{\mathcal{M}}_{as}^{lk}(\mathbf k)\overline{\mathcal{M}}_{bt}^{nq}(\mathbf k'),
\end{align}
with $\overline{\mathcal M}_{as}^{lk}$ the moments of the distribution function defined in \cref{eq:gyromomentsfinal}.

Finally, the result in \cref{eq:collophl} is used in \cref{eq:collopmoments} in order to find the Hermite-Laguerre moments $C_{ab}^{pj}$ of the full Coulomb collision operator expressed in \cref{eq:JiCabgyrok}.
This yields
\begin{equation}
    C_{ab}^{pj}=\sum_{l,k,n,q=0}^\infty\sum_{m=0}^k\sum_{r=0}^q \frac{L_{km}^l L_{qr}^n}{\sqrt{2^p p!}} C_{ab,lkm}^{pj,nqr},
\label{eq:hermitelaguerrecpj2}
\end{equation}
with
\begin{align}
    C_{ab,lkm}^{pj,nqr}(\mathbf k, \mathbf k') &=\sum_{u=0}^{min(2,l,n)}\sum_{i=0}^{min(l,n)-u} \sum_{d=-l-n+2i+2u}^{l+n-2i-2u}\sum_{z{,p,p'}=0}^\infty D_{abuidz{pp'}}^{lkmnqr}(\mathbf k, \mathbf k') I,
\label{eq:collmomentsfinal}
\end{align}
and
\begin{equation}
    I = \int f_{aM} P_{l+n-2i-2u}^d(\overline v_\parallel /\overline v){\nu_{*abu}^{lm,nr}(\overline v^2)} H_p(\overline s_{\parallel a})L_j(\overline s_{\perp a}^2) {\overline s_{\perp a}^{d+2z}} {L_p^z(\overline s_{\perp a}^2)L_{p'}^{z+d}(\overline s_{\perp a}^2)} \frac{B_\parallel^*}{m_a}d \overline v_\parallel d \overline \mu.
\end{equation}
The integral factor $I$ can be performed analytically by first rewriting the product of two Laguerre polynomials as a single one using
\begin{equation}
    L_r^m(x)L_j(x)=\sum_{s=0}^{r+j}\overline d_{rjs}^m L_s(x),
\end{equation}
with
\begin{equation}
    \overline d^m_{rjs}=\sum_{r_1=0}^r \sum_{j_1=0}^j \sum_{s_1=0}^s L_{r r_1}^{-1/2} L_{j j_1}^{m-1/2} L_{s s_1}^{-1/2} (r_1+j_1+s_1)!,
\end{equation}
expressing the resulting Hermite-Laguerre basis in terms of Legendre-Associated Laguerre using \cref{eq:plljhpljminus1}, and writing the phase-space volume $(B_\parallel^{*}/m)d \overline v_\parallel d \overline \mu$ as $\overline v^2 d\overline v d \overline \xi$ with $\overline \xi = \overline v_\parallel/\overline v$.
This yields
\begin{equation}
    I = \sum_{{h}=0}^{{p+}z+j}{\sum_{g=0}^{g+p'}}\sum_{s=0}^{p+2g}\sum_{t=0}^{g+\floor{p/2}}{d_{pjh}^{std}\overline d_{p'hg}^{z+d}} \left(T^{-1}\right)^{std}_{pg} C_{*abu}^{st,lm,nr}\frac{(s+d)!}{(s-d)!}\frac{\delta_{l+n-2i-2u,s}}{4\pi(s+1/2)}.
\label{eq:finalI}
\end{equation}
\citet{Ji2009} present an analytical closed expression ready to be numerically implemented of the factor $C_{*abu}^{st,lm,nr}=\int f_{Ma}\nu_{*abu}^{lm,nr}(v^2)L_{t}^{s+1/2}(v^2) v^s d\mathbf v$.
We note that the long-wavelength limit can be found by setting $d=z=0$ in the collision operator \cref{eq:collmomentsfinal}.
This yields the Hermite-Laguerre moments of the collision operator moments found in \citet{Jorge2017}.

\section{Small Mass{-}Ratio Approximation}
\label{sec:smallmassratio}

In this section, we simplify the electron-ion and the ion-electron collision operator in \cref{eq:coulombop} by taking advantage of the small electron-to-ion mass ratio $m_e/m_i$, and derive their expressions in the gyrokinetic regime.
We first consider the electron-ion collision operator.

In $(\mathbf x, \mathbf v)$ phase-space coordinates, the electron-ion Coulomb collision operator can be greatly simplified by taking advantage of the fact that the ion thermal {speed}, is small in comparison to the electron thermal {speed}, for $T_e \sim T_i$.
To first order in $m_e/m_i$, the electron-ion Coulomb collision operator, also called  Lorentz pitch-angle scattering operator, can be written as \citep{Helander2002}
\begin{equation}
    \begin{split}
        C_{ei}&=\frac{n_i L_{ei}}{v_{the}^3}\frac{\partial}{\partial \mathbf c_e}\cdot\l[\frac{1}{c_e}\frac{\partial f_e}{\partial \mathbf c_e}-\frac{\mathbf c_e}{c_e^3}\l(\mathbf c_e \cdot \frac{\partial f_e}{\partial \mathbf c_e}\r)\r],
    \end{split}
    \label{eq:cei0}
\end{equation}
with $\mathbf c_e = \mathbf v/v_{the}$.
We expand $f_e$ according to \cref{eq:faexp}.
We note that the expansion in \cref{eq:faexp} is an eigenbasis of the pitch-angle scattering operator $C_{ei}$ with eigenvalue $l(l+1)$ \citep{Ji2008}.
Therefore, we write
\begin{equation}
    C_{ei}=-f_{eM}\sum_{l,k}\frac{n_i L_{ei}}{v_{the}^3 c_e^3}\frac{l(l+1)}{\sqrt{\sigma_k^l}}L_k^{l+1/2}\left(c_e^2\right) \mathbf Y^l(\mathbf c_e) \cdot {\mathbf M}_e^{lk}(\mathbf x).
    \label{eq:cei0eig}
\end{equation}
We now Fourier transform the moments $\mathbf M_e^{lk}$ in \cref{eq:cei0eig} as $\mathbf M_e^{lk}(\overline{\mathbf R}) = \int \mathbf M_e^{lk}(\mathbf k)e^{i \mathbf k \cdot \overline{\mathbf R}} d \mathbf k$ and write the gyroaveraged collision operator $C_{ei}$ as
\begin{equation}
    \left<C_{ei}\right>_{\overline{\mathbf R}}=-\int d\mathbf k e^{i \mathbf k \cdot \overline{\mathbf R}} f_{eM}\sum_{l,k}\frac{n_i L_{ei}}{v_{the}^3 c_e^3}\frac{l(l+1)}{\sqrt{\sigma_k^l}} L_k^{l+1/2}\left(c_e^2\right) \left<\mathbf Y^l(\mathbf c_e) e^{i \mathbf k \cdot \mathbf \rho_e}\right>_{\overline{\mathbf R}}\cdot \mathbf M_{e}^{lk}(\mathbf k).
    \label{eq:cei0eiggyro}
\end{equation}
Using the Jacobi-Anger expansion of \cref{eq:jacoan}, \cref{eq:bessLeg}, the inverse basis transformation \cref{eq:plljhpljminus1}, and the identities $J_{-p}(x)=(-1)^p J_p(x)$ and
\begin{equation}
    L_r^m(x)=\sum_{j=0}^{r}\binom{m+r-j-1}{r-j}L_j(x),
\end{equation}
we obtain
\begin{align}
    \left<\mathbf Y^l(\mathbf v) e^{i \mathbf k \cdot \mathbf \rho_e}\right>_{\mathbf R} &= \sum_{m=-l}^l \sum_{r=0}^\infty \sum_{j=0}^{r}\sum_{s=0}^{2j}\sum_{t=0}^j \sqrt{\frac{\pi^{1/2}l!}{2^l(l-1/2)!}\frac{(l-m)!}{(l+m)!}}\frac{i^m \mathbf e^{lm}}{(m+r)!}\frac{(m+r-j-1)!}{(r-j)!(m-1)!}\nonumber\\
    &\times v_{the}^l (T^{-1})_{0j}^{stm}b_e^{2r+m}e^{-b_e^2}c_e^{l+s}P_l^m(\cos \varphi)P_s^m(\cos \varphi) L_t^{s+1/2}(c_e^2),
\label{eq:Ylexpgyro}
\end{align}
with $b_e=k_\perp \rho_{the}/2$.
The collision operator in \cref{eq:cei0eiggyro} represents the gyrokinetic electron-ion collision operator.

Equation (\ref{eq:Ylexpgyro}) provides an expression of the pitch-angle scattering operator $\left<C_{ei}\right>$ in \cref{eq:cei0eiggyro} suitable for projection onto a Hermite-Laguerre basis, i.e.
\begin{align}
    C_{ei}^{pj}&=\int \left<C_{ei}\right> \frac{H_p\left(\frac{\overline v_{\parallel}}{v_{tha}}\right)L_j\left(\frac{\overline \mu B}{T_a}\right) }{\sqrt{2^p p!}} \frac{B_\parallel^*}{m_a} d\overline v_\parallel d \overline \mu d \overline \theta=2\pi\sum_{l=0}^{p+2j}\sum_{k=0}^{j+\floor{p/2}}\frac{(T^{-1})_{pj}^{lk0}v_{the}^3}{\sqrt{2^p p!}} I_{ei}^{lk},
\label{eq:ceiopjgg}
\end{align}
where we define
\begin{equation}
    I_{ei}^{lk}=\int \left<C_{ei}\right> c_e^l P_l(\cos \varphi) L_k^{l+1/2}(c_e^2) c_e^2 dc_e d\cos \varphi.
\label{eq:iei0lk}
\end{equation}
An analytical form for the integral factor $I_{ei}^{lk}$ can be derived using the expression for $\left<C_{ei}\right>$, \cref{eq:cei0eiggyro}, and \cref{eq:Ylexpgyro}, yielding
\begin{align}
    I_{ei}^{lk}(\mathbf k)&=-\sum_{u,v}\frac{n_e n_i L_{ei}}{v_{the}^{6-u}}\frac{u(u+1)}{\pi \sqrt{\sigma_v^u}} \sum_{m=-u}^u\mathbf M_{e}^{lk}(\mathbf k) \cdot \mathbf e^{um} \sum_{r=0}^\infty \sum_{i=0}^{r}\sum_{s=0}^{2i}\sum_{t=0}^i(T^{-1})_{0i}^{stm}e^{-b_e^2}\nonumber\\
    &\times \sqrt{\frac{u!}{2^u(u-1/2)!}\frac{(u-m)!}{(u+m)!}}\frac{i^m  b_e^{2r+m}}{(m+r)!}\frac{(m+r-i-1)!}{(r-i)!(m-1)!}I_{L k t}^{l s u v}I_{Pm}^{l u s},
\end{align}
with $I_{L k t}^{l s u v}$ and $I_{Pm}^{l u s}$ defined by
\begin{equation}
    I_{L k t}^{l s u v}=\int L_k^{l+1/2}(x)L_t^{s+1/2}(x)e^{-x} x^{(l+u+v)/2-1}dx,
\end{equation}
and
\begin{equation}
    I_{Pm}^{l u s}=\int_{-1}^{1} P_l(x)P_{u}^m(x)P_{s}^m(x) \frac{dx}{2},
\end{equation}
respectively.
The electron fluid moments $\mathbf M_{e}^{lk}$ can be cast in terms of Hermite-Laguerre moments $\overline N_e^{lk}$ using the expressions in Eqs. (\ref{eq:gcmalnfinal}), (\ref{eq:gyromomentsstar}), and (\ref{eq:gyromomentsfinal}).
The factor $I_{L k t}^{l s u v}$ can be analytically evaluated by expanding the associated Laguerre polynomials using \cref{eq:asslaguerre}, which leads to
\begin{equation}
    I_{L k t}^{l s u v}=\sum_{m_1=0}^{k}\sum_{m_2=0}^{t}{L_{km_1}^{l}L_{tm_2}^s}(m_1+m_2+(l+u+v)/2-1)!.
\end{equation}
Similarly, the factor integral $I_{Pm}^{l u s}$ can be calculated using an extended version of Gaunt's formula \citep{Gaunt1929}, yielding \citep{Mavromatis1999}
\begin{align}
    I_{Pm}^{l u s}&=(-1)^m\tj l u s 0 0 0 \tj l u s 0 m {-m} \sqrt{\frac{(s+m)!(u+m)!}{(s-m)!(u-m)!}}.
\label{eq:gaunt}
\end{align}
We note that, in \cref{eq:gaunt}, the Wigner 3-j symbol $\tj {j_1} {j_2} {j_3} {m_1} {m_2} {m_3}$ is related to the Clebsch-Gordan coefficients $\left<j_1 m_1 j_2 m_2 | j_3 m_3 \right>$ via \citep{Olver2010}
\begin{equation}
    \tj {j_1} {j_2} {j_3} {m_1} {m_2} {m_3}=\frac{(-1)^{j_1-j_2-m_3}}{\sqrt{2 j_3+1}}\left<j_1 m_1 j_2 m_2 | j_3 (- m_3) \right>,
\end{equation}
with the Clebsch-Gordan coefficients given by
\begin{align}
    &\left<j_1 m_1 j_2 m_2 | j_3 m_3 \right>=\delta_{m_3,m_1+m_2}\sqrt{\frac{(2j_3+1)(j_3+j_1-j_2)!(j_3-j_1+j_2)!(j_1+j_2-j_3)!}{(j_1+j_2+j_3+1)!}}\nonumber\\
    &\times\sqrt{(j_3+m_3)!(j_3-m3)!(j_1-m_1)!(j_1+m_1)!(j_2-m_2)!(j_3-m_3)!}\nonumber\\
    &\times\sum_k \frac{(-1)^k}{k!(j_1+j_2-j_3-k)!(j_1-m_1-k)!(j_2+m_2-k)!}\nonumber\\
    &\times\frac{1}{(j_3-j_2+m_1+k)!(j_3-j_1-m_2+k)!},
\label{eq:cgcoeffs}
\end{align}
where the summation in \cref{eq:cgcoeffs} is extended over all integers $k$ that make every factorial in the sum nonnegative \citep{Bohm1993}.


We now turn to the ion-electron collision operator $C_{ie}$.
To first order in $m_e/m_i$, this is given by \citep{Helander2002}
\begin{equation}
    \begin{split}
        C_{ie}&=
        \nu_{ei}\frac{m_e}{m_i}\frac{\partial}{\partial \mathbf v}\cdot\left(\mathbf v f_i
        +\frac{T_e}{m_i}
        \frac{\partial f_i}{\partial \mathbf v} 
        \right),
    \end{split}
    \label{eq:cie}
\end{equation}
where the electron-ion friction force is neglected for simplicity.
%
We simplify \cref{eq:cie} by using \cref{eq:orderingftildei}, therefore approximating the distribution function $f_i$ by its gyroaveraged component $f_i \simeq \left<\overline F_i\right>_{\overline{\mathbf R}}$, and retaining the lowest-order terms in the $\epsilon_\delta$ expansion.
This allows us to convert the $C_{ie}$ operator in \cref{eq:cie} to the gyrocenter variables $\overline{\mathbf Z}$ using the chain rule at lowest order in $\epsilon_\delta$, i.e. to express \cref{eq:cie} in $\mathbf Z$ coordinates using the guiding-center transformation in \cref{eq:gcv,eq:gcmu,eq:gcx} and approximate $\mathbf Z \simeq \overline{\mathbf Z}$.
The velocity derivatives can be written as a function of $\overline{\mathbf Z}$ using the chain rule, yielding
\begin{equation}
\begin{split}
    \frac{\partial \left<\overline F_i\right>_{\overline{\mathbf R}}}{\partial \mathbf v}&=\mathbf b \frac{\partial \left<\overline F_i\right>_{\overline{\mathbf R}}}{\partial \overline v_\parallel}+\mathbf c\left( \sqrt{\frac{2 m_a \overline \mu}{B}}\frac{\partial \left<\overline F_i\right>_{\overline{\mathbf R}}}{\partial \overline \mu}- \frac{1}{\Omega_i}\mathbf a \cdot \nabla_{\overline{\mathbf R}}\left<\overline F_i\right>_{\overline{\mathbf R}}\right),
\end{split}
\end{equation}
where we define $\mathbf c=(\cos \overline \theta \mathbf e_1 + \sin \overline \theta \mathbf e_2)$ and $\mathbf a = \mathbf c \times \mathbf b = (-\sin \overline \theta \mathbf e_1 + \cos \overline \theta \mathbf e_2)$.
The ion-electron collision operator can therefore be written as
\begin{align}
    C_{ie}&=\nu_{ei}\frac{m_e}{m_i}\left[3\left<\overline F_i\right>_{\overline{\mathbf R}}+v_\parallel \frac{\partial \left<\overline F_i\right>_{\overline{\mathbf R}}}{\partial \overline v_\parallel}+2\overline \mu \frac{\partial \left<\overline F_i\right>_{\overline{\mathbf R}}}{\partial \overline \mu}-\sqrt{\frac{2 B \overline \mu}{m_i}}\frac{\mathbf a \cdot \nabla_{\overline{\mathbf R}}}{\Omega_i}\left<\overline F_i\right>_{\overline{\mathbf R}}\right.\nonumber\\
    &+\frac{T_e}{m_i}\left(\frac{\partial^2 \left<\overline F_i\right>_{\overline{\mathbf R}}}{\partial \overline v_\parallel^2}+\frac{2 m_i \overline \mu}{B}\frac{\partial^2 \left<\overline F_i\right>_{\overline{\mathbf R}}}{\partial \overline \mu^2}+\frac{\mathbf a \cdot \nabla_{\overline{\mathbf R}} \mathbf a \cdot \nabla_{\overline{\mathbf R}} }{\Omega_i^2}\left<\overline F_i\right>_{\overline{\mathbf R}}\right.\nonumber\\
    &\left.\left.-2\sqrt{\frac{2 m_i \overline \mu}{B}}\mathbf a \cdot \nabla_{\overline{\mathbf R}} \frac{\partial \left<\overline F_i\right>_{\overline{\mathbf R}}}{\partial \overline \mu}+\frac{m_i}{B}\frac{\partial \left< \overline F_i \right >_{\overline{\mathbf R}}}{\partial \overline \mu}\right)\right].
\end{align}
We now Fourier transform both $T_e$ and $\left<\overline F_i\right>_{\overline{\mathbf R}}$ and gyroaverage $C_{ie}$, yielding
\begin{align}
        \left<C_{ie}\right>_{\overline{\mathbf R}}&=\nu_{ei}\frac{m_e}{m_i}\int e^{i (\mathbf k + \mathbf k')\cdot \mathbf R} \left[\left<\overline F_i\right>_{\overline{\mathbf R}}+\frac{\partial \left<\overline F_i\right>_{\overline{\mathbf R}}}{\partial \overline v_\parallel}+2\overline \mu \frac{\partial \left<\overline F_i\right>_{\overline{\mathbf R}}}{\partial \overline \mu}\right.\nonumber\\
        &+J_0(k_\perp' \overline{\rho_i})\frac{T_e(\mathbf k')}{m_i}\left(\frac{\partial^2 \left<\overline F_i\right>_{\overline{\mathbf R}}}{\partial \overline v_\parallel^2}+\frac{2 m_i \overline \mu}{B}\frac{\partial^2 \left<\overline F_i\right>_{\overline{\mathbf R}}}{\partial \overline \mu^2}+\frac{m_i}{B}\frac{\partial \left<\overline F_i\right>_{\overline{\mathbf R}}}{\partial \overline \mu}\right)\nonumber.\\
        &\left.+\frac{T_e(\mathbf k')}{m_i}i\frac{\left<\overline F_i\right>_{\overline{\mathbf R}}}{2\Omega_i^2} \left[J_0(k_\perp' \overline{\rho_i})k_\perp^2+J_2(k_\perp' \overline{\rho_i})\mathbf k \mathbf k : \left(\mathbf e_1 \mathbf e_1-\mathbf e_2 \mathbf e_2\right)\right]\right.\nonumber\\
        &\left.-\frac{T_e(\mathbf k')}{m_i}\mathbf k \cdot \mathbf e_2 i J_1(k_\perp' \overline{\rho_i})\frac{2 m_i \overline v_\perp}{B \Omega_i}\frac{\partial \left<\overline F_i\right>_{\overline{\mathbf R}}}{\partial \overline \mu}\right],
\label{eq:cie12}
\end{align}
where we have used the identities $\left< \mathbf a e^{i \mathbf k' \cdot \mathbf \overline{\rho_i}}\right>_{\overline{\mathbf R}}=i J_1(k_\perp' \overline{\rho_i}) \mathbf e_2$ and $\left< \mathbf a \mathbf a e^{i \mathbf k' \cdot \mathbf \overline{\rho_i}}\right>_{\overline{\mathbf R}}=(1/2)[J_0(k_\perp' \overline{\rho_i})\left(\mathbf e_1 \mathbf e_1+\mathbf e_2 \mathbf e_2\right)+J_2(k_\perp' \overline{\rho_i})\left(\mathbf e_1 \mathbf e_1-\mathbf e_2 \mathbf e_2\right)]$.
Finally, we take Hermite-Laguerre moments of the gyroaveraged ion-electron collision operator $\left<C_{ie}\right>_{\overline{\mathbf R}}$ in \cref{eq:cie12}, using the expansion of $ \left<\overline F_i\right>_{\overline{\mathbf R}}$ in \cref{eq:gyrof}, yielding
\begin{align}
    C_{ie}^{pj}&=\nu_{ei}\frac{m_e}{m_i}\int e^{i (\mathbf k + \mathbf k')\cdot \mathbf R}\sum_{l,k}\left[A_{lk}^{pj}+e^{-b_i^2}\frac{T_e(k_\perp')}{T_i}\sum_{r=0}^{\infty}\frac{b_i^{2r}}{r!}\left(\sum_{s=0}^{r+j}d_{rjs}^0 B_{lkrs}^{pj}\right.\right.\nonumber\\
    &\left.\left.+\sum_{v=0}^{r}\sum_{s=0}^{v+j+1}\frac{d_{vjs}^1 i \rho_{thi}^2 b_i^2\delta_{l p}\delta_{k s} }{4(r+1)(r+2)}\mathbf k \mathbf k : \left(\mathbf e_1 \mathbf e_1-\mathbf e_2 \mathbf e_2\right)\right)\right]\overline N_i^{lk}(\mathbf k),
\label{eq:ciepj}
\end{align}
with $A_{lk}^{pj}$ given by
\begin{equation}
\begin{split}
    A_{lk}^{pj}&=2j\delta_{lp}\delta_{kj-1}-(p+2j)\delta_{lp}\delta_{kj}-\sqrt{ p (p-1)}\delta_{l p-2}\delta_{kj},
\end{split}
\end{equation}
and $B_{lkrs}^{pj}$ by
\begin{equation}
    \begin{split}
        B_{lkrs}^{pj}&=\sqrt{p(p-1)}\delta_{l p-2}\delta_{ks}+\frac{T_i}{m_i}\frac{i k_\perp^2}{2 \Omega_i^2}-\sum_{i=0}^{s-1}(3+2s)\delta_{lp}\delta_{ki}+\sum_{i=0}^{s-2}2s\delta_{lp}\delta_{ki}\\
        &+i \mathbf k \cdot \mathbf e_2 \frac{2 v_{thi}}{ \Omega_i}\frac{b_i}{r+1}[(1+s)\delta_{l p}\delta_{k s}-s\delta_{l p}\delta_{k s-1}].
    \end{split}
\end{equation}

\section{Conclusion}

In this work, a formulation of the nonlinear gyrokinetic Coulomb collision operator is derived, providing an extension of a previously derived nonlinear Coulomb drift-kinetic collision operator to the gyrokinetic regime.
This constitutes a key element necessary to perform quantitative studies of turbulence, flows, and, in general, of the plasma dynamics in the periphery of magnetized fusion devices.
The gyroaveraged collision operator is cast in terms of parallel and perpendicular velocity integrals of the gyroaveraged distribution function at arbitrary $k_\perp \rho_s$, yielding the formula in \cref{eq:JiCabgyrok}.
In order to provide an analytical formulation of the Coulomb collision operator ready to be used in pseudospectral formulations of the gyrokinetic equation for distribution functions arbitrarily far from equilibrium and for an arbitrary collisionality regime, the Hermite-Laguerre moments of the gyroaveraged collision operator are evaluated, yielding \cref{eq:hermitelaguerrecpj2}.
Furthermore, the electron-to-ion mass ratio is used to simplify the form of the electron-ion and ion-electron collision operators, yielding \cref{eq:ceiopjgg} and \cref{eq:ciepj}, respectively.

{We conclude by noting that the present collision operator is derived by porting the Coulomb operator to the gyrocenter phase-space by using a framework valid up to second order in the expansion parameter $\epsilon$, yielding second order accurate momentum and energy conservation laws. The use of the techniques developed here to analytically gyroaverage the Coulomb operator and obtain its projection onto an orthogonal polynomial basis should, in principle, be applicable to collision operators of the Fokker-Planck type that add the necessary correction terms in order to ensure exact conservation laws \citep{Brizard2004,Sugama2015,Burby2015}.}

\section{Acknowledgements}

We thank L. M. Perrone for the helpful insight on the relation between spherical basis tensors and spherical harmonics {and the anonymous referees for their careful review of our article.}
This work has been carried out within the framework of the EUROfusion Consortium and has received funding from the Euratom research and training programme 2014-2018 and 2019-2020 under grant agreement No 633053, from Portuguese FCT (Fundação para a Ciência e Tecnologia) under grant PD/BD/105979/2014, carried out as part of the training in the framework of the Advanced Program in Plasma Science and Engineering (APPLAuSE,) sponsored by FCT under grant No. PD/00505/2012 at Instituto Superior Técnico, and from the Swiss National Science Foundation.
The views and opinions expressed herein do not necessarily reflect those of the European Commission.

\appendix
\addtocontents{toc}{\protect\setcounter{tocdepth}{0}}

\section{Basis Tensors}
\label{app:basistensors}

In this appendix, we derive the form of the basis tensors $\mathbf e^{lm}$ used in the definition of $\mathbf Y^l(\mathbf v)$ in \cref{eq:ylvfull}.
We start with the $l=1$ case, for which \cref{eq:ylvfull} yields
\begin{equation}
    \mathbf Y^1(\mathbf v) = \mathbf v = \sqrt{\frac{4 \pi}{3}}v \sum_{m=-1}^1 Y_{1m}(\phi,\theta) \mathbf e^{1m}.
\label{eq:Y1defe}
\end{equation}
The spherical basis vectors $\mathbf e^{1m}$ can then be derived from \cref{eq:Y1defe} by expressing the vector $\mathbf v$ in spherical coordinates as
\begin{equation}
    \mathbf v = v\left(\sin \phi \cos \theta \mathbf e_x+\sin \phi \sin \theta \mathbf e_y+\cos \phi \mathbf e_z\right),
\end{equation}
and using the identities for the spherical harmonics
\begin{equation}
    Y_{1m}(\phi,\theta)=
\begin{cases}
    \sqrt{\frac{3}{8 \pi}}\sin \phi e^{-i \theta}, &m=-1,\\
    \sqrt{\frac{3}{4 \pi}}\cos \phi, &m=0,\\
    -\sqrt{\frac{3}{8 \pi}}\sin \phi e^{i \theta}, &m=1,\\
\end{cases}
\end{equation}
therefore obtaining
\begin{equation}
    \mathbf e^{1m}=
\begin{cases}
    \frac{\mathbf e_x-i \mathbf e_y}{\sqrt{2}}, &m=-1,\\
    \mathbf e_z, &m=0,\\
    -\frac{\mathbf e_x+i \mathbf e_y}{\sqrt{2}}, &m=1.\\
\end{cases}
\label{eq:buvectorse1m}
\end{equation}

We now construct spherical basis tensors $\mathbf e^{lm}$ from the spherical basis vectors $\mathbf e^{1m}$ leveraging the techniques developed for the angular momentum formalism in quantum mechanics \citep{Zettili2009,Snider2018}.
As a first step, we note that the basis vectors $\mathbf e^{1m}$ are eigenvectors of the angular momentum matrix $G_z$
\begin{equation}
    G_z=i
  \begin{pmatrix}
    0 & -1 & 0\\
    1 &  0 & 0\\
    0 &  0 & 0
  \end{pmatrix},
\end{equation}
with eigenvalue $m$, that is
\begin{equation}
    G_z \cdot \mathbf e^{1m} = m \mathbf e^{1m}.
\end{equation}
As a second step, we note that the relationship between the basis vectors $\mathbf e_\alpha$ for $\alpha=(x,y,z)$ and the angular momentum matrices $G_\alpha$ is given by
\begin{equation}
    G_{\alpha} = -i \mathbf e_{\alpha} \cdot \epsilon,
\label{eq:gnmatr}
\end{equation}
with $\epsilon$ the standard Levi-Civita tensor.
In index notation, \cref{eq:gnmatr} can be written as
\begin{equation}
    \left({G_{\alpha}}\right)_{kl}=-i\sum_{j=1}^3\left(e_{\alpha}\right)_j \epsilon_{jkl}.
\end{equation}
The raising $G_+$ and lowering $G_-$ operators (corresponding to the ladder operators in quantum mechanics), defined by
\begin{equation}
    G_{\pm}=G_x \pm i G_y.
\end{equation}
allows us to obtain the basis vectors $\mathbf e^{1\pm1}$ from $\mathbf e^{10}$ using
\begin{equation}
    G_{\pm}\mathbf e^{10}=\mathbf e^{1\pm1}.
\end{equation}
In addition, we have that
\begin{equation}
    \mathbf e^{1 -1}=(G_{-})^2 \mathbf e^{11}.
\end{equation}

We can now define the spherical tensor basis $\mathbf e^{l m}$ that define the irreducible tensors $\mathbf Y^{l}$.
We start with the spherical basis tensor
\begin{equation}
    \mathbf e^{ll}=\mathbf e^{11}\mathbf e^{11}...\mathbf e^{11},
\end{equation}
formed by the product of $l$ basis vectors $\mathbf e^{11}$.
Similarly to $\mathbf Y^l(\mathbf v)$, this tensor is of rank $l$, symmetric, and totally traceless, as $\mathbf e^{11}\cdot \mathbf e^{11} = 0$.
Furthermore, we note that $\mathbf e^{ll}$ is an eigenvector with eigenvalue $l$ of the angular momentum tensor $G_z^{l}$, with $G_n^{l}$ a tensor of rank $2l$ defined by
\begin{equation}
\begin{split}
    \left[G_\alpha^{l}\right]_{a_1 a_2 ... a_l b_1 b_2 ... b_l}=\sum_{j'k'...l'}&\left\{\left[G_\alpha\right]_{a_1 b_1}\delta_{a_2 b_2}...\delta_{a_l b_l}+ \delta_{a_1 b_1}\left[G_\alpha\right]_{a_2 b_2}...\delta_{a_l b_l}\right.\\
    &\left.+...+\delta_{a_1 b_1}\delta_{a_2 b_2}...\left[G_\alpha\right]_{a_l b_l}\right\}.
\end{split}
\end{equation}
The remaining basis tensor elements $\mathbf e^{l m}$ can be obtained by applying the tensorial lowering operator $G^l_-=G_x^{l}-i G_y^{l}$ to $\mathbf e^{ll}$, namely
\begin{equation}
    \mathbf e^{l m} = \sqrt{\frac{(l+m)!}{(2l)!(l-m)!}}\left(G^{l}_-\cdot^{l}\right)^{l-m}\mathbf e^{ll},
\label{eq:elmbasistensor}
\end{equation}
with $m=-l,-l+1,...,-1,0,1,...,l$ and $\left(G^{l}_-\cdot^{l}\right)^{l-m}\mathbf e^{ll}$ a tensor of order $l$ built by the application of the $G^{l}_-\cdot^{l}$ operator to $\mathbf e^{ll}$ $l-m$ times.
The normalization factor in \cref{eq:elmbasistensor} is obtained by requiring that the contravariant $\mathbf e^{lm}$ and the covariant $\mathbf e^l_m$ basis tensors form an orthonormal basis, i.e.
\begin{equation}
    \mathbf e^{l m} \cdot \mathbf e^{l}_{m'} = \delta_{m,m'}.
\label{eq:elmconst}
\end{equation}
In order to find a covariant basis $e^l_m$, we start with the case $l=1$ and note that the set of vectors $\mathbf e^{1}_m=(\mathbf e^{1m})^* = (-1)^m \mathbf e^{1-m}$, with $(\mathbf e^{1}_m)^*$ the complex conjugate of $\mathbf e^{1}_m$ satisfies \cref{eq:elmconst}.
We therefore define $\mathbf e^l_m=(\mathbf e^{lm})^*$, and use \cref{eq:elmconst} to normalize $\mathbf e^{lm}$.
For computational purposes, we note that the tensor $\mathbf e^{l m}$ can also be written as a function of the basis vectors $\mathbf e^{1m}$ as \citep{Snider2018}
\begin{equation}
    \mathbf e^{l m}=N_{lm}\sum_{n=0}^{\floor{\frac{l+m}{2}}}a_n^{lm}\left\{(\mathbf e^{11})^{m+n}(\mathbf e^{1-1})^{n}(\mathbf e^{10})^{l-m-2n}\right\}_{TS},
\end{equation}
where $N_{lm}=\sqrt{(l+m)!(l-m)!2^{l-m}/(2l)!}$ and $a_n^{lm}=l!/[2^n n!(m+n)!(l-m-2n)!]$.

\section{Basis Transformation}
\label{app:tlkpj1}

In this section, we derive a closed-form expression for the $T_{lkm}^{pj}$ and $(T^{-1})^{lkm}_{pj}$ coefficients defined in \cref{eq:plljhplj,eq:plljhpljminus1}.
By multiplying \cref{eq:plljhplj} by a Hermite and a Laguerre polynomial and by an exponential of the form $e^{-\overline v^2}$, and integrating over the whole $\overline v_\parallel$ and $\overline \mu$ space, we obtain the following integral expression for $T_{lkm}^{pj}$
\begin{equation}
    T_{lkm}^{pj}=\frac{v_{tha}^{m-l}}{2^p p! \sqrt{\pi}} \int \frac{\overline v^l}{\overline v_\perp^m} P_l^m\left(\frac{\overline v_\parallel}{\overline v}\right) L_k^{l+1/2}\left(\frac{\overline v^2}{v_{tha}^2}\right) H_p\left(\frac{\overline v_{\parallel a}}{v_{tha}}\right)L_j\left(\frac{\overline v_\perp^2}{v_{tha}^2}\right) e^{-\frac{v^2}{v_{tha}^2}}\frac{d \mathbf v}{2 \pi}.
\label{eq:appbas1}
\end{equation}
We first write the integrand in \cref{eq:appbas1} in terms of $\overline \xi=\overline v_\parallel/\overline v$ and $\overline v$ coordinates using the basis transformation in \cref{eq:plljhpljminus1}, yielding
\begin{equation}
\begin{split}
     T_{lkm}^{pj}&= \sum_{l'=0}^{p+2j}\sum_{k'=0}^{j+\floor{p/2}}\frac{(l+1/2)k!}{(l+k+1/2)!}T_{l'k'}^{pj}\\
     &\times\int_{-1}^1 \frac{P_l^m(\overline \xi) P_{l'}(\overline \xi)}{(1-\overline \xi)^2} d \overline \xi \int_0^\infty x_a^{(l+l'-m+1)/2}L_k^{l+1/2}(x_a)L_{k'}^{l'+1/2}(x_a) dx_a,
\end{split}
\label{eq:appbas2}
\end{equation}
where we used the fact that $(T^{-1})_{lk}^{pj}=T_{lk}^{pj}\sqrt{\pi}2^p p! k! (l+1/2)/(k+l+1/2)!$ \citep{Jorge2017}.
The $\xi$ integral in \cref{eq:appbas2} is performed by expanding $P_{l}$ as a finite sum of the form
\begin{equation}
    P_{l}(x)=\sum_{s=0}^l c_s^l x^s,
\end{equation}
with the coefficients $c_s^l=2^l[(l+s-1)/2]!/[s!(l-s)!((s-l-1)/2)!]$, and using the relation between associated Legendre functions $P_l^m(x)$ and Legendre polynomials $P_l(x)$
\begin{equation}
    P_l^m(x) = (-1)^m(1-x^2)^{m/2}\frac{d^m P_l(x)}{dx^m}.
\end{equation}
The $x$ integral in \cref{eq:appbas2} is performed by using the expansion of the associated Laguerre polynomials in \cref{eq:asslaguerre}.
The $T_{lkm}^{pj}$ coefficient can then be written as
\begin{align}
    T_{lkm}^{pj}&=\sum_{l'=0}^{p+2j}\sum_{k'=0}^{j+\floor{p/2}} T_{l'k'}^{pj}\frac{(l'+1/2)k'!}{(l'+k'+1/2)!} \sum_{m_1=0}^k\sum_{m_2=0}^{k'}\sum_{s_1=0}^l\sum_{s_2=0}^{l'}L_{k m_1}^l L_{k' m_2}^{l'}\nonumber\\
    &\times \frac{c_{s_1}^l c_{s_2}^{l'}}{2} \frac{s_1!}{(s_1-m)!} \frac{\left[1+(-1)^{s_1+s_2-m}\right]}{s_1+s_2+1-m}\left(m_1+m_2+\frac{l+l'-m+1}{2}\right)!.
\end{align}
The inverse transformation coefficients $(T^{-1})^{lkm}_{pj}$ defined by \cref{eq:plljhpljminus1} can be found similarly, yielding
\begin{equation}
    (T^{-1})^{lkm}_{pj}=\frac{2^p p! \sqrt{\pi} k! (l+1/2)(l-m)!}{(k+l+1/2)!(l+m)!}T_{lkm}^{pj}.
\end{equation}

\bibliographystyle{jpp}
\bibliography{library}

\end{document}